\documentclass[a4paper,twocolumn,11pt]{quantumarticle}
\pdfoutput=1
\usepackage{amsmath}
\usepackage{amssymb}
\usepackage{graphicx}
\usepackage{bbm}
\usepackage[colorlinks=true,urlcolor=blue,menucolor=blue,citecolor=blue,linkcolor=blue]{hyperref}
\usepackage{xcolor}
\renewcommand{\vec}[1]{\mathbf{#1}}
\newcommand{\ket}[1]{|{#1}\rangle}
\newcommand{\bra}[1]{\langle{#1}|}

\newcommand{\tr}{\mathrm{Tr}}

\makeatletter
\def\myfnsymbol#1{\expandafter\@alph2\csname c@#1\endcsname}
\def\@myfnsymbol#1{\ensuremath{\ifcase#1\or \dagger\or \ddagger\or
   \mathsection\or \mathparagraph\or \|\or **\or \dagger\dagger
   \or \ddagger\ddagger \else\@ctrerr\fi}}
\let\@fnsymbol\@myfnsymbol
\makeatother

\begin{document}
\title{Optimal Quantum State Tomography with Noisy Gates}

\author{Violeta N.~Ivanova-Rohling$^*$}
\email{violeta.ivanova-rohling@uni-konstanz.de}
\affiliation{Department of Physics, University of Konstanz, D-78457 Konstanz, Germany}
\affiliation{Zukunftskolleg, University of Konstanz, D-78457 Konstanz, Germany}
\affiliation{%
Department of Mathematical Foundations of Computer Sciences, Institute of Mathematics and Informatics, Bulgarian Academy of Sciences,
Akad.\ G.\ Bonchev, block 8,
1113 Sofia, Bulgaria\\$^*$These authors contributed equally to this work.}
\author{Niklas Rohling$^*$}
\email{niklas.rohling@uni-konstanz.de}
\affiliation{Department of Physics, University of Konstanz, D-78457 Konstanz, Germany}
\author{Guido Burkard}
\email{guido.burkard@uni-konstanz.de}
\affiliation{Department of Physics, University of Konstanz, D-78457 Konstanz, Germany}

\begin{abstract}
Quantum state tomography (QST) represents an essential tool for the characterization, verification, and validation (QCVV) of quantum processors.
Only for a few idealized scenarios, there are analytic results for the optimal measurement set for QST. E.g.,
in a setting of non-degenerate measurements, an optimal minimal set of measurement operators for QST has eigenbases which are mutually unbiased.
However, in other set-ups, dependent on the rank of the projection operators and the size of the quantum system, the optimal choice of measurements for efficient QST needs to be numerically approximated.
We have generalized this problem by introducing the framework of \emph{customized efficient QST}.
Here we extend customized QST and look for the optimal measurement set for QST in the case where some of the quantum gates applied in the measurement process are noisy.
To achieve this, we use two distinct noise models: first, the depolarizing channel, and second, over- and under-rotation in single-qubit and to two-qubit gates (for further information, please see Methods).
We demonstrate the benefit of using entangling gates for the efficient QST measurement schemes for two qubits at realistic noise levels, by comparing the fidelity of reconstruction of our optimized QST measurement set to the state-of-the-art scheme using only product bases. 
\end{abstract}

\maketitle
\section{Introduction}
\subsection{Background}
In the past decades, the mounting evidence that quantum algorithms can solve specific tasks with efficiency beyond the capability of a state-of-the-art classical computer has led to considerable interest in the field of quantum computing.
A major turning point was Shor's algorithm for prime factorization \cite{Shor}.
In addition, hard optimization problems are expected to be efficiently solved on a quantum device with potentially enormous consequences for multiple fields. Feynman’s proposal to use quantum computers for the efficient simulation of quantum systems for which classical simulation is hard \cite{QuantumSimulationsRMP}represents another high-impact application.
Various physical hardware platforms are being developed for quantum computation 
\cite{ionsreview,Aruteetal2019,supercondqubitsreview,LossDiVincenzo,KloeffelLoss2013,NV-centers}.

The increasing size and complexity of quantum devices call for more sophisticated techniques for calibration, certification, and evaluation of their performance. 
The field of quantum characterization, verification, and validation (QCVV) offers various state-of-the-art protocols and techniques to evaluate the performance of a quantum system.
Quantum state tomography (QST) \cite{QSTbookchapter2004}, a prominent QCVV technique, allows for the reconstruction of a given quantum state from measurement data.
Others include quantum process tomography, randomized benchmarking (RB) \cite{Emersonetal2005,Mavadiaetal2018}, and gate set tomography \cite{Merkeletal2013,Blume-Kohoutetal2013,Mavadiaetal2018,Nielsenetal2020}. 

While QST is known as the ``gold standard" for the verification of a quantum device \cite{Crameretal2010}, as it provides comprehensive information for a given quantum state, its computational costs make it infeasible for a system larger than few qubits.
Moreover, full QST can be time-consuming even if performed on  small systems, say building blocks of a quantum computer of only one or two qubits.
Therefore, the search for efficient measurement schemes for QST is of high practical importance. 

Optimal QST measurement schemes are known for specific ideal and noise-free scenarios.
For a $d$-dimensional Hilbert space, the ideal choice is a set of $d+1$ measurement operators whose eigenbases are mutually unbiased bases (MUBs) \cite{WoottersFields89}. 
For generalized measurements, using ancillary systems, symmetric, informationally complete positive operator-valued measures (SIC-POVMs) are optimal \cite{Rehaceketal2004,Renesetal}.
For a situation where one out of $N$ qubits is measured, an optimal quorum \cite{BodmannHaas2018} consists of projectors on so-called mutually unbiased subspaces.
Numerically optimized QST measurement sets consisting of independent rank-1 projection operators \cite{VioletaNiklas} and projectors on half-dimensional subspaces in dimension six \cite{VioletaGuidoNiklas} have been obtained.
In the first case, the numerical solution outperforms a set that constitutes projectors from a set of MUBs and in the latter case, the solution approximates mutually unbiased subspaces.

When implementing QST on real systems, one is inevitably confronted with the presence of noise and decoherence during every quantum operation.
Despite the importance of QST  as an established tool for determining the state of a quantum system, rigorous and systematic research on optimizing QST with noisy gates is lacking.
 Limited research into optimal  measurement schemes for QST in the presence of noise for single- and two-qubit systems exist
 \cite{deBurgh2008,Mohammadi2014,Miranowicz2014}, however, in these works fine-tuning of noisy entangling gates are not  considered. In \cite{Miranowicz2014}, the measurement set of a photonic 2-qubit quantum system is optimized to be maximally robust to a general measurement error. In \cite{Mohammadi2014}, the general question of QST under measurement constraints is investigated, with an implementation on a single (photonic) qubit. In \cite{deBurgh2008} a QST on a single qubit is performed with a set of generalized (possibly overcomplete) measurements.
\subsection{A framework for optimal quantum state tomography in noisy systems}
\begin{figure}
    \centering
    \includegraphics[width=\columnwidth]{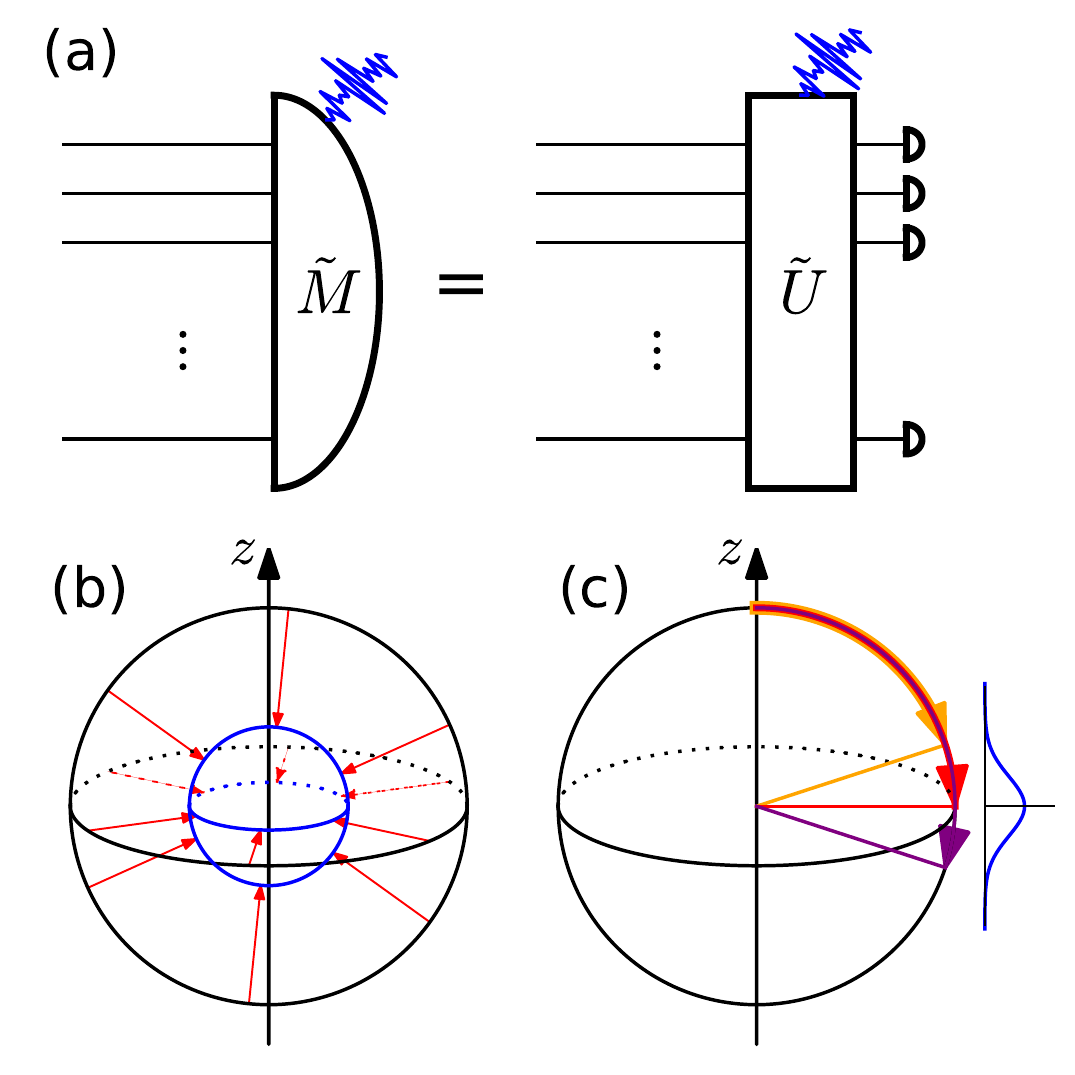}
    \caption{(a): Noisy measurement $\tilde M$ realized by a noisy unitary operation $\tilde U$ followed by a noiseless measurement in the standard basis.
    (b,c): Noise models illustrated here for a single qubit on the Bloch sphere.
    (b) The depolarizing channel shrinks the Bloch vector.
    (c) Over- and under-rotation describes errors during a quantum gate where the rotation angle fluctuates according to a Gaussian distribution. The example shows an intended  $\pi/2$-rotation about an axis in the $xy$-plane.}
    \label{fig:one}
\end{figure}

Here, we extend the framework by looking for optimal QST schemes in noisy systems, and modify the QST quality measure defined by  Wootters and Fields \cite{WoottersFields89}, see Sec.~\ref{sec:effect_of_noise}. 
They expressed the information about a quantum state obtained by performing measurements on this state as
\begin{equation}
I = - \ln\frac{V}{V_0} - \mathrm{const.}
\end{equation}
where $V_0$ is the volume of all possible quantum states. The confidence volume $V$  is defined as the volume of the rectangular parallelepiped which includes the part of the distribution, assumed to be Gaussian, with probability density larger than $1/e$ times its maximum.
For dimension $d$, $V$ can be expressed as 
\begin{equation}
    V = \frac{V_1\cdots V_{d{+}1}}{\mathcal{Q}}.
\end{equation}
Here, $V_j$ is the confidence volume of the $(d{-}1)$-dimensional subspace spanned by the projectors $P_{jk}$ on eigenstates $k=1,\ldots,d$ of the measurement $j$
and $\mathcal{Q}$ is the geometric quality measure given by the volume of the rectangular parallelepiped spanned by $\{P_{jk}|j{=}1,\ldots,d{+}1;k{=}1,\ldots,d{-}1\}$.
The number of repetitions $N_j$ and the probabilities of the measurement outcomes, $p_{jk} = \operatorname{Tr}(P_{jk}\rho)$ are related to $V_j$ by
\begin{equation}
    V_j \sim \frac{1}{N_j^{(d{-}1)/2}} \sqrt{p_{j1}\cdots p_{jd}}.
    \label{eq:Vj}
\end{equation}
The averaged information gain reads $\langle I\rangle = \int d\mu_d(\rho) I(\rho)$, where $\mu_d(\rho)$ is the Haar measure of the density matrices for a $d$-dimensional Hilbert space.
Importantly, $\langle I\rangle$ depends on the choice of the QST measurement set only via $\mathcal{Q}$ because $\langle \ln (\operatorname{Tr}[P \rho])\rangle$ is the same constant for any projector $P$ of the same rank.
Thus $\mathcal{Q}$ can be used as a quality measure for a QST measurement set.

We consider measurements that are realized by first applying a sequence of quantum gates followed by a measurement in a standard basis, see Fig~\ref{fig:one} (a).
The quality of the selected QST measurement scheme then depends on the choice of gates and how much the chosen gates are affected by noise.
There are multiple noise models for quantum gates, and, for simplicity, we focus on two of those models.
Note, however, that our approach is not limited to any noise model.
First, we consider a Hamiltonian of the form $H=\lambda P_\lambda$ where we -- for now -- consider $P_\lambda=\ket{\lambda}\bra{\lambda}$ to be projector of rank 1.
Switching this Hamiltonian on for a time $t$ yields the unitary
$$U(\phi)=\mathbbm{1}-P_\lambda + e^{i\phi} P_\lambda$$
where we have defined $\phi:=\lambda t$.
To model noisy gate operations, we assume that $\phi$ cannot be controlled precisely by the applied pulses but instead follows a Gaussian distribution $p_N(\phi)$ with $\langle\phi\rangle = \phi_0$ and a standard deviation of $\sigma = \sqrt{\langle(\phi-\phi_0)^2\rangle}$, see Fig.~\ref{fig:one} (c).
We assume that variance scales linearly with $\phi_0$, $\sigma^2=2r\phi_0$.
Representing a density matrix $\rho$ in the eigenbasis of $U(\phi)$ and averaging over the Gaussian, reveals that the off-diagonal elements of $\int d\phi\, p_N(\phi)U^\dag\rho U$ which involve $\ket{\lambda}$ or $\bra{\lambda}$ are modified by $e^{-r\phi_0}e^{-i\phi_0}$, i.e. additionally to the desired phase factor, they  decay exponentially.
Although we consider this model for two-qubit gates, later on, we use the term over- and under-rotation \cite{Devittetal2013,Reineretal2018} for it.
Second, we consider the depolarizing channel \cite{NielsenChuang}, see Fig.~\ref{fig:one} (b).
If noise affects any quantum state in the same way, based on the waiting time, independent of whether  a quantum gate is applied during this time or not, all the projectors which describe one measurement are modified in the same way.
However, the waiting time depends on the overall time of the quantum gate sequence needed for the measurement.

The structure and contents of this article are as follows.
We consider a single-qubit system, see Sec.~\ref{sec:single-qubit}, assuming that measurements along the $z$ axis of the Bloch sphere and rotations around the $z$ axis are error-free while rotations around an axis in the $xy$ plane are noisy.
As in  \cite{TrieuThesis}, the above two noise models coincide for the single qubit. 
Then, we turn to a two-qubit system, see Sec.~\ref{two-qubits}, with error-free single-qubit gates and entangling two-qubit gates affected by noise.
While in general, any non-trivial interaction can lead to an entangling gate, we explicitly consider the Heisenberg interaction, see Sec.~\ref{sec:Heisenberg}, and the Ising interaction, see Sec.~\ref{sec:Ising}.
These interactions or interactions which yield equivalent gates are relevant for certain quantum computing platforms, namely, the Heisenberg interaction is present for spin qubits in semiconductor quantum dots \cite{LossDiVincenzo} and an Ising-equivalent interaction for resonant gates \cite{Huang,Chowetal2011}.
Our methods for solving the resulting numerical optimization problem are explained in Sec.~\ref{sec:numerics} and our results are presented in Secs.~\ref{sec:resultsexploratoryanalysis} and \ref{sec:resultsMUBstart}.
There, we investigate the change in the quality of the optimal measurement schemes for different noise levels as well as the corresponding normalized times for the entangling gates.
We perform numerical simulations to demonstrate the benefit of using entangling gates for efficient QST by comparing the fidelity of reconstruction obtained by our measurement schemes to the fidelity of reconstruction using nine product bases and show superiority of using entangling gates for noise level corresponding to average gate fidelities between values around $0.8$ and $0.9$ depending on the noise model, see Sec.~\ref{sec:resultsfidelityofreconstruction}.
At the same time, we quantify the advantage of using a QST measurement set optimized for the specific noise level over the use of the QST scheme optimal for the noiseless case. 
Importantly, to overcome the limitations imposed by the considered noise models, we validate the advantages of using entangling gates for QST on a real quantum device, running QST on both, an actual quantum device and a simulator emulating it.
The results of the experiments on the device confirm that in certain scenarios, a better fidelity of the QST procedure is obtained using entangling gates in comparison to the state-of-the-art product bases.  
We conclude in Sec.~\ref{sec:conclusions}.
\subsection{Summary of contributions}
Here, we summarize the significant contributions and assumptions of our paper.
Most importantly, we introduce a \textit{formal evaluation scheme for the efficiency of QST quorums in noisy settings}. Note that a such tool has not been available before, although any experimentally implemented QST deals with noise.

Secondly, we apply our scheme to multiple relevant scenarios, including a single-qubit and a two-qubit system with noisy entangling gates, where two important noise models (depolarizing channel and over- and under-rotation) were implemented. We consider two-qubit gates based on Heisenberg and Ising interaction. To make our investigations relevant for practitioners, we identify realistic noise levels by comparison to recent literature.

Furthermore, we formulate the goal of finding the optimal QST quorum under noise as an optimization problem and \textit{derive a mathematically meaningful quality measure} for each respective interaction and noise model and perform an \textit{extensive exploratory analysis} using global optimization approaches  for exploration combined with local approaches for exploitation. Based on this analysis, it was that the best solutions have the specific structure of three measurement bases of product states and two bases with highly entangled gates related to the standard set of MUBs (where the entangled states are maximally entangled); we derive an analytical expression for the optimized solutions for the depolarizing channel.

Moreover, using simulations, we compare the fidelity of reconstruction for QST with MUBs and our numerically optimized QST to QST with nine separable bases which represents the current state-of-the-art approach for QST under noise.
As an important result, we find a \textit{significant advantage in using entangling gates} compared to the nine separable bases for noise levels already achieved in experiments.  
Finally, for validation, we compare the performance of QST under noise using entangling gates and product bases on a real quantum computer.
The results confirm that for certain realistic scenarios it is more advantageous to use entangling gates for QST.
Additionally, we find a crossover behavior where QST using nine product bases as measurement bases becomes the more advantageous approach with an increasing number of experimental runs of each measurement.

\section{Results}
\subsection{Quantum state tomography with noise}
\label{sec:effect_of_noise}
We now investigate the performance of quantum state tomography quorums under the influence of noise.
Noise which is independent of the choice of the measurement will only lead to a constant change in $\langle I\rangle$.
This can be compensated by an increased number of experimental runs, but the optimal QST measurement set remains the same.
We consider the more interesting situation where the noise depends on the choice of measurements.
We will analyze how the noise affects the averaged information gain $\langle I\rangle$ and incorporate this dependence by modifying the quality measure $\mathcal{Q}_N$.

We consider the case where the measurement $j$ is described by a POVM, $\{F_{jk}\}$ with $\sum_k F_{jk}=\mathbbm{1}$.
In the noise-free case, $F_{jk}=P_{jk}$ where $P_{jk}$ are projection operators of rank $l_{jk}$
with $P_{jk}P_{ji}=\delta_{ik}P_{jk}$ and $\sum_k{l_{jk}}=d$.
We look at  operators being affected by noise such that
\begin{equation}
F_{jk} = q_{j}\left(P_{jk}-\frac{l_{jk}\mathbbm{1}}{d}\right) + \frac{l_{jk}\mathbbm{1}}{d}.
\label{eq:proj-noise}
\end{equation}
Here, the traceless part of the operator is rescaled with a factor of $q_{j}\le1$.
Then the outcomes of the noise-affected measurement $j$ follow a multinomial distribution with probabilities $p'_{jk}=\tr(F_{jk}\rho)=q_{j}p_{jk}+l_{jk}(1-q_{j})/d$.
From $p'_{jk}$ we can calculate the probabilities in the noiseless case
\begin{equation}
    p_{jk}=\frac{p'_{jk}}{q_{j}} - \frac{l_{jk}(1-q_{j})}{q_{j}d}.
\end{equation}
As in \cite{WoottersFields89}, we assume that the multinomial distribution is well approximated by a Gaussian.
We then use \eqref{eq:Vj} with rescaled probabilities $p_{jk}'/q_{j}$ and the relation \eqref{eq:proj-noise} to calculate the confidence intervals as
\begin{equation}
   V_j \sim \sqrt{N_j}\prod_k \frac{1}{\sqrt{N_j}}\sqrt{ p_{jk}+\frac{l_{jk}(1-q_{j})}{q_{j}d} }.
\end{equation}
Rescaling is necessary as the noise-affected measurement outcome is related to the outcome of a corresponding noise-free measurement not only by a shift in probability but also by a factor of $q_j$
In order to include the effect of the noise, described here by the value of $q_{j}$, we need to compute
$\langle \ln\left(p_{jk}+{l_{jk}(1-q_{j})}/(q_{j}d)\right)\rangle$
using the Haar measure $\mu_d(\rho)$.
We will calculate this expression for the cases $d=2$ and $d=4$, as they are  investigated in detail in this paper.
For $d=2$, we can calculate this expression analytically by integrating over the Bloch sphere in cylindrical coordinates for $P=\ket{0}\bra{0}$,
\begin{equation}
\begin{split}
    \left\langle \ln\left(p_{jk}{+}c\right)\right\rangle
     = & -\frac{5}{6} +2c(1{+}c)+c^2(3{+}2c)\ln(c)\\
       & -(1{+}c)^2(2c{-}1)\ln(1{+}c),
\end{split}
\end{equation}
which we can expand for $c\propto 1-q_{j}\ll1$ up to linear order in $(1-q_{j})$ yielding
\begin{equation}
    \left\langle \ln\left(p_{jk}+\frac{1-q_{j}}{2q_{j}}\right)\right\rangle \approx -\frac{5}{6} + \frac{3}{2}(1-q_{j}),
\end{equation}
where $l_{jk}=1$ for $d=2$.
For $d=4$, we compute the average numerically.
We obtain random density matrices using the representation
$\rho=UDU^\dag$, see \cite{EisertPlenio}, where $D$ is a diagonal matrix with the eigenvalues of $\rho$ on the diagonal obtained as the differences between the elements of $\{0,r_1,r_2,r_3,1\}$ where $\{r_1,r_2,r_3\}$ is an ordered set of three values taken from a uniform distribution over the interval $[0,1]$.
We generate random unitary matrices $U$ using the approach described in ~\cite{FMezzadri2007}.
We use linear regression to fit the data  to get an estimate of the coefficient. We use $10^7$ matrices, projectors on the four basis vectors, and 40 data points on the interval $q_j\in[0.9,1]$.
We obtain for $l_{jk}=1$,
\begin{equation}
    \left\langle \ln\left(p_{jk}+\frac{1-q_{j}}{4q_{j}}\right)\right\rangle \approx {\rm const.} + 1.195(1-q_{j}).
    \label{eq:averaging-4d-expval}
\end{equation}
We disregard terms of higher than linear order in $1-q_j$ and set $N_j=N$, then we obtain
\begin{equation}
    \langle I \rangle = \ln\mathcal{Q} +\frac{d^2-1}{2}\ln{N} - s\sum_j (1-q_j) + const.
\end{equation}
where $s$ depends on the rank of the projectors involved in the measurements, for $d=2$, we have $s=3/2$, for non-degenerate measurements in 4D, we have $s=2.39$.
The $q_j$-dependent part of the averaged information gain $\langle I\rangle$ will now be included in a modified quality measure $\mathcal{Q}_N$,
$\mathcal{Q}_N = \mathcal{Q}\prod_j[\exp(q_j{-}1)]^s$.
The average information gain depends on $\mathcal{Q}_N$ by $\langle I\rangle = \ln\mathcal{Q}_N  +\frac{d^2-1}{2}\ln{N} + const.$
In linear order in $1{-}q_j$, $\mathcal{Q}_N$ simplifies to
\begin{equation}
    \mathcal{Q}_N = \mathcal{Q}\prod_j(q_j)^s.
    \label{eq:quality-measure-with-noise}
\end{equation}
We will use $\mathcal{Q}_N$ in order to find optimal QST measurement set under noisy conditions.

\subsection{Single qubit}
\label{sec:single-qubit}
We consider a qubit and assume that while the standard measurement in the $z$ basis is free of errors, rotations about any axis in the $xy$ plane are error-prone.
We assume that the angle of such rotations follows the Gaussian distribution described above, $p(\theta) \sim \exp((\theta-\theta_0)^2/(2\sqrt{2}r \theta_0))$.
So, if we want to project onto the state $\cos(\theta_0/2)|0\rangle + \sin(\theta_0/2)e^{i\phi}|1\rangle$, we actually project onto the mixed state
\[
\rho = \frac{1}{2}\left(\begin{array}{cc}1 & 0 \\0 & 1\end{array}\right)
       +\frac{e^{-r\theta_0}}{2}\left(\begin{array}{cc}\cos\theta_0 & e^{-i\phi}\sin\theta_0 \\e^{i\phi}\sin\theta_0 & -\cos\theta_0\end{array}\right).
\]
We see that the traceless part of the density matrix decays exponentially with increasing $0<\theta_0<\pi$, i.e. the imperfection of the quantum gate yields a depolarization channel for the qubit.
The traceless parts of the density matrix can be written as a three-component real vector, the Bloch vector $\vec n$, such that $\rho=\mathbbm{1}/2 + \vec n \cdot \boldsymbol\sigma /2$ where $\boldsymbol\sigma$ is a vector of Pauli matrices.
In the specific situation considered here the length of $\vec n$ depends on the angle $\theta$,
\[
\vec n = e^{-r|\theta|}\left(\begin{array}{c} \sin(\theta)\cos(\phi)\\
                                            \sin(\theta)\sin(\phi)\\
                                            \cos(\theta)
                         \end{array}\right).
\]
In other words, the value of $q_j$ defined above is given by $\exp(-r\theta_j)$ as it is for the depolarizing channel.
Numerically one finds that the maximal $\mathcal{Q}_N$ is three Bloch vectors all with the same $\theta$ and the phases can be chosen to be $\phi_1=0$, $\phi_2=2\pi/3$, and $\phi_3=4\pi/3$.
Then, the quality measure with the exponent $s=3/2$, see Sec.~\ref{sec:effect_of_noise}, reads
\[
\mathcal{Q}_N = \frac{3\sqrt{3}}{2} e^{-9r|\theta|/2}\cos(\theta)\sin^2(\theta),
\]
and the optimal angle is $\theta=\arctan(\sqrt{81r^2/16+2}-9r/4)$.

\subsection{Two qubits with noisy entangling gates}
\label{two-qubits}
We formalize the question ``When does it make sense to include entangling gates in a QST measurement scheme?"
for non-degenerate measurements.

Each of the measurements included in a QST quorum is assumed to be carried out by first applying a unitary operation $U_j$
where $j=1,\ldots,5$ for non-degenerate measurements 
 to the unknown quantum state and then performing a measurement in the standard basis $\{|00\rangle,|01\rangle,|10\rangle,|11\rangle\}$.
The task is now to find $U_1,\ldots,U_5$ which yield the highest $\mathcal{Q}_N$.

\subsubsection{Universal quantum gates with noisy entangler}
\label{sec:universal2-q-gate}
General unitary operators acting on two qubits can be represented by
\begin{equation}
    U_j = U^{q1}(\boldsymbol\Phi_{j,1}^{(1)})
          U^{q2}(\boldsymbol\Phi_{j,1}^{(2)})
          U^{tq}(\boldsymbol\beta_j)
          U^{q1}(\boldsymbol\Phi_{j,2}^{(1)})
          U^{q2}(\boldsymbol\Phi_{j,2}^{(2)})
\label{eq:universal2-q-gate}
\end{equation}
where $U^{qk}(\boldsymbol\Phi^{(k)}_{jm})$ ($k,m=1,2$; $j=1,\ldots,5$) is a local one-qubit gate applied to qubit $k$ defined by the three real parameters
$\boldsymbol\Phi=(\phi,\psi,\chi)$,
\begin{equation}
    U^{qk}(\boldsymbol\Phi) = \left(
    \begin{array}{cc}
    \cos(\phi)e^{i\psi} & \sin(\phi)e^{i\chi}\\
    -\sin(\phi)e^{-i\chi} & \cos(\phi)e^{-i\psi}
   \end{array}\right).
\end{equation}
The gate $U^{tq}(\boldsymbol\beta_j)$ is a universal two-qubit gate, i.e., together with the local gates any desired $U_j\in SU(4)$ can be realized.
As a parametrization of $U^{tq}(\boldsymbol\beta_j)$, the Hamiltonian
\begin{equation}
    H_p(\boldsymbol\beta_j) = \sum_{\alpha=x,y,z}\beta_{j\alpha} \sigma_\alpha^{(1)}\otimes\sigma_\alpha^{(2)}
\end{equation}
where $\sigma_\alpha^{(k)}$ are the Pauli matrices for the qubit $k$, can be used \cite{KrausCirac2001,Khanejaetal2001,Zhangetal2003},
\begin{equation}
    U^{tq}(\boldsymbol\beta_j) = \exp\left(-iH_p(\boldsymbol\beta_j)\right).
    \label{eq:2qg-representation}
\end{equation}
The operators $H_p$ and $U^{tq}$ are diagonal in the Bell basis
$\{|\Phi_+\rangle, |\Psi_+\rangle, |\Phi_-\rangle, |\Psi_-\rangle\}$,
where $|\Phi_\pm\rangle = (|00\rangle\pm|11\rangle)/\sqrt{2}$, 
$|\Psi_\pm\rangle = (|01\rangle\pm|10\rangle)/\sqrt{2}$
with the corresponding eigenvalues
$\eta_{00}^j = \beta_{jx} - \beta_{jy} + \beta_{jz}$,
$\eta_{01}^j = \beta_{jx} + \beta_{jy} - \beta_{jz}$,
$\eta_{10}^j = -\beta_{jx} + \beta_{jy} + \beta_{jz}$,
$\eta_{11}^j = -\beta_{jx} - \beta_{jy} - \beta_{jz}$
for $H_p$ and $e^{-i\eta_{ik}^j}$ for $U^{tq}$.
While, any interaction which yields a universal two-qubit gate can be applied to reproduce a desired $U^{tq}$, the representation in Eq.~(\ref{eq:2qg-representation}) provides a very convenient way to do this for any Hamiltonian which is of the form $H_p$ with any fixed values of $\boldsymbol\beta_j$
including the Heisenberg ($\beta_{jx}=\beta_{jy}=\beta_{jz}$) and the Ising interactions ($\beta_{jx}=\beta_{jy}=0$),
Any $U^{tq}$ can be generated by applying any of these gates \textit{three} times with appropriate single-qubit gates in between. 

\subsubsection{Heisenberg interaction}
\label{sec:Heisenberg}
Now, we demonstrate how to compose the universal two-qubit gate for the situation where the entangling gate originates from the Heisenberg exchange interaction ($\beta_{jx}=\beta_{jy}=\beta_{jz}$),
\begin{equation}
 H_{\text ex}=\lambda_H |\Psi_-\rangle\langle \Psi_-|
\end{equation}
where $\lambda_H$ is the interaction strength.
This situation can be realized by spin qubits in semiconductor quantum dots \cite{LossDiVincenzo}.
Following the work by Fan \textit{et al.} \cite{Fanetal2005}, $U^{tq}$ is then defined by three parameters
$\boldsymbol\alpha_j=(\alpha_{j,1},\alpha_{j,2},\alpha_{j,3})$, which are directly related to the amount of time the Heisenberg interaction is switched on,
$$\alpha_{jk}=\frac{(t_{jk}^{\rm end} -t_{jk}^{\rm start})\lambda_H}{\pi},$$
yielding SWAP$^{\alpha_{jk}}$ gates.
We assume that the interaction is switched on and off instantaneously.
Thus the $\alpha_{jk}$ can be considered as normalized entangling times with the unit $\pi/\lambda_H$, $|\boldsymbol\alpha_j|_1=\sum_k \alpha_{jk}$ is then the normalized entangling time for the measurement $j$ and additionally, we define $\alpha:=\sum_j|\boldsymbol\alpha_j|_1$ as the normalized overall entangling time applied within the gate sequences for a QST measurement set.
The gate $U^{tq}$ is composed as \cite{Fanetal2005}
\begin{equation}
\begin{split}
    U^{tq}(\boldsymbol\alpha_j) =&
    \sigma_z^{(1)}\sigma_x^{(2)}{\rm SWAP}^{\alpha_{j,1}}
    \sigma_z^{(1)}{\rm SWAP}^{\alpha_{j,2}}\\
    &\times\sigma_x^{(2)}{\rm SWAP}^{\alpha_{j,3}}.
    \end{split}
\end{equation}
In the -- now unconventionally sorted -- Bell basis
$\{|\Psi_+\rangle, |\Phi_+\rangle, |\Phi_-\rangle, |\Psi_-\rangle\}$,
$U^{tq}$ is given by
\begin{equation}
    U^{tq}(\boldsymbol\alpha_j) =
    \operatorname{diag}(1, e^{i\alpha_{j,1}\pi}, e^{i\alpha_{j,2}\pi}, e^{i\alpha_{j,3}\pi}).
\end{equation}
Switching on the Heisenberg interaction creates an entangling gate. In our model, the normalized time of the Heisenberg interaction between two spin qubits determines how much the noise affects the quantum system. Since, aside from the QST optimal measurement scheme, we are interested in the role the entangling gate plays in the optimal solution, we evaluate the normalized times $\alpha_{jk}$ during which the Heisenberg interaction is switched on.

\subsubsection{Ising interaction}
\label{sec:Ising}
We construct the universal two-qubit gate using the Ising interaction
\begin{equation}
    H_{\rm Ising} = \lambda_I \sigma_z^{(1)}\otimes\sigma_z^{(2)}.
\end{equation}
We note that we can reproduce the representation in Eq.~(\ref{eq:2qg-representation}) by the sequence
\begin{equation}
\label{eq:Ising2qgate}
    U^{tq}(\boldsymbol\beta_j) = \prod_{k=x,y,z}
    (U^{(1)}_k)^\dag
    (U^{(2)}_k)^\dag
    e^{-i\beta_{jk} \sigma_z^{(1)}\otimes\sigma_z^{(2)}}
    U^{(2)}_k
    U^{(1)}_k
\end{equation}
with the single-qubit operations
$U_x^{(m)}=\exp(-i\pi\sigma_y^{(m)}/4)$,
$U_y^{(m)}=\exp(i\pi\sigma_x^{(m)}/4)$,
$U_z^{(m)}=\mathbbm{1}$ ($m=1,2)$.
As for the Heisenberg interaction, we define normalized entangling times for each switching on of the interaction
$$\tau^I_{jk} = \frac{(t_{jk}^{\rm end}-t_{jk}^{\rm start})\lambda_I}{\pi},$$
for the gate sequence used for realizing measurement number $j$,
$\tau_j^I = \sum_k\tau_{jk}^I$,
and for all sequences within one QST measurement set,
$\tau^I=\sum_j\tau^I_j$.
The unit of these \textit{times} is again $\pi/\lambda_I$, i.e., $\pi$ divided by interaction strength.
Note that, also interactions of the form $H\sim(a_x\sigma_x^{(1)}+a_y\sigma_y^{(1)})\otimes\sigma_z^{(2)}$
yield a gate  locally equivalent to the gate provided by the Ising interactions.
Those interactions are effectively present when resonant gates are applied \cite{Chowetal2011}.

\subsection{Results from the exploratory analysis: geometric measure and normalized entangling time}
\label{sec:resultsexploratoryanalysis}
\subsubsection{Optimization via multiple runs of local search}
\begin{figure}
    \centering
    \includegraphics[width=\columnwidth]{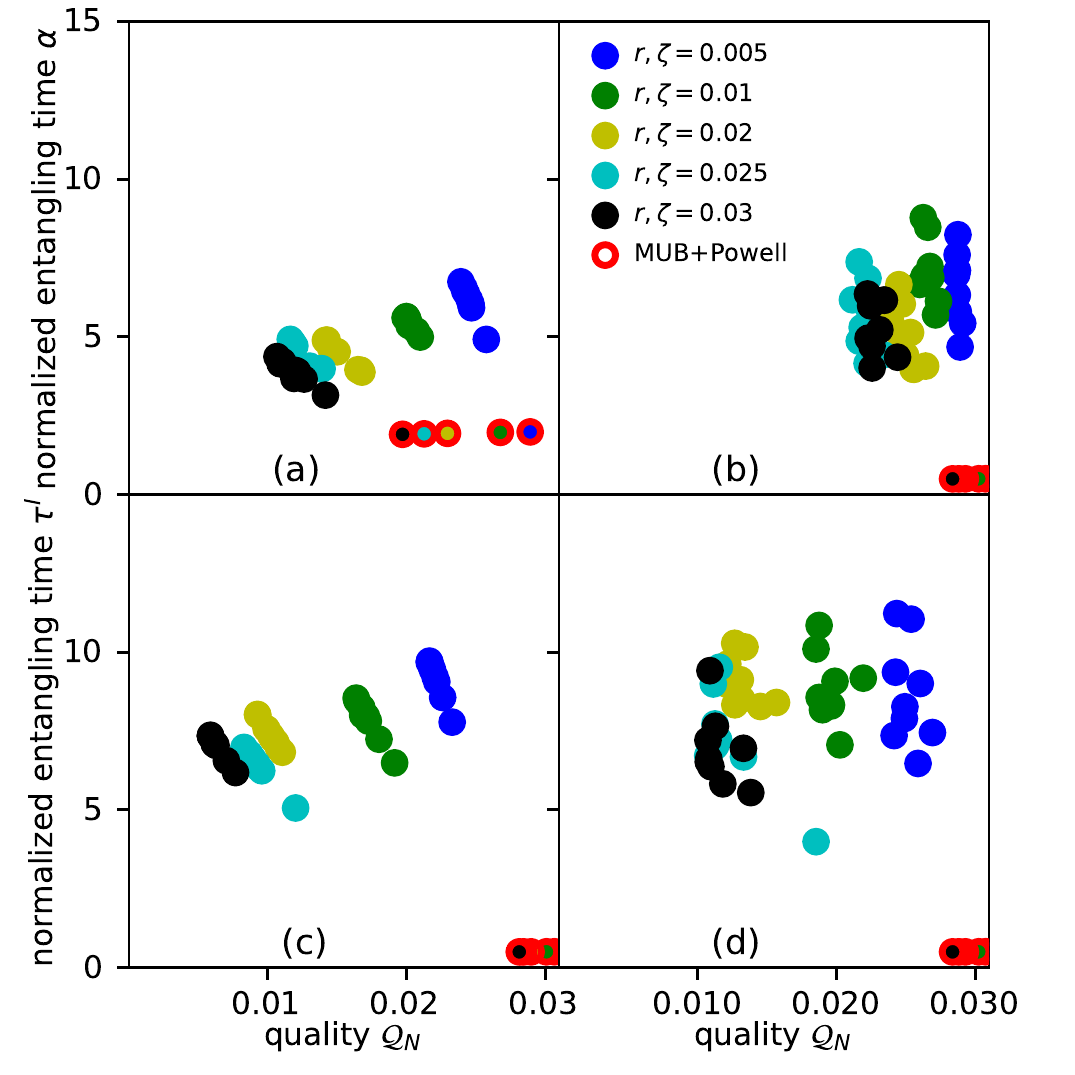}
    \caption{The quality and entanglement time of selected solutions (sets of quorums) discovered by  using Powell's method with 500 random diverse starting  points,  for the Heisenberg (a,b) and the Ising interaction (c,d) with noise described by the depolarizing channel (a,c) or over- and under-rotation (b,d). Only the 10 best solutions are visualized for each noise level. 
    The points with the red circles represent the optimized quorums resulting from using Powell's method with standard MUBs as starting points being as good as or better than the best results from the exploratory analysis. Here $r$ and $\zeta$ are the noise parameters for the over-and-under rotation noise model and for the depolarizing channel, respectively.
  }
    \label{fig:exploratorylocal}
\end{figure}

One approach to look for an optimal solution is to use parallel searches with many well chosen starting points in order to explore well the space of potential optimal solutions. This approach has been used successfully for similar problems in \cite{VioletaNiklas,VioletaGuidoNiklas}. We use Powell's method as a local search with a set of 500 diverse starting points, each at least 0.01 distance from each other using a Jaccard-based distance measure. For more details, see section \ref{sec:numerics}.
In Fig.~\ref{fig:exploratorylocal} we present the quality of a selection of the quorums discovered by the above approach.

\subsubsection{Global Optimization via Simulated Annealing}
In Fig.~\ref{fig:exploratoryglobal}, results of the global optimization via Simulated Annealing are presented. Ten runs are performed for each noise level.
The resulting solutions are contrasted to the optimized solution discovered by Powell's method using the set of MUBs, see Sec.~\ref{sec:standardMUBstartingpoint}, as starting points for the local search.
In principle, there are infinitely many possible sets of MUBs that can be used. For our comparison, we use a set of MUBs that is known to have a shorter entangling times $\tau^I=1/2$ and $\alpha=2$
than other known choices. In the following, we refer to this set as the standard set of MUBs.
The results of the exploratory analysis demonstrate that high-quality solutions are discovered with entanglement close to or below that of the standard set of MUBs. Here $\zeta$ is the noise parameter for the depolarizing channel with Ising and Heisenberg interactions and $r$ is the noise parameter for the over-and-underrotation channel with the two types of interactions (see Method section for the precise definition). For higher levels of noise, e.g. $\zeta$ and $r$ take values around $3\%$, the solutions discovered by the global optimization approach  are close to the ones found by using a local search with the standard MUBs as starting points of the search. This, as well as the consideration that the a full set of MUBs represents an ideal quorum of projectors in the absence of noise, motivated us to use the aforementioned set of MUBs as a starting point of a local search for lower levels of noise and compare the results to the results from the global optimization.
The relation that a shorter entangling time corresponds to a higher quality measure for optimized quorums does not hold for over- and under-rotation with Ising interaction due to special invariants, discussed below.

\subsubsection{Invariance of the $\mathcal{Q}_N$ for specific applications of the Ising interaction with over- and under-rotation}
\label{sec:isingoverunderinvar}
\begin{figure}
    \centering
    \includegraphics[width=\columnwidth]{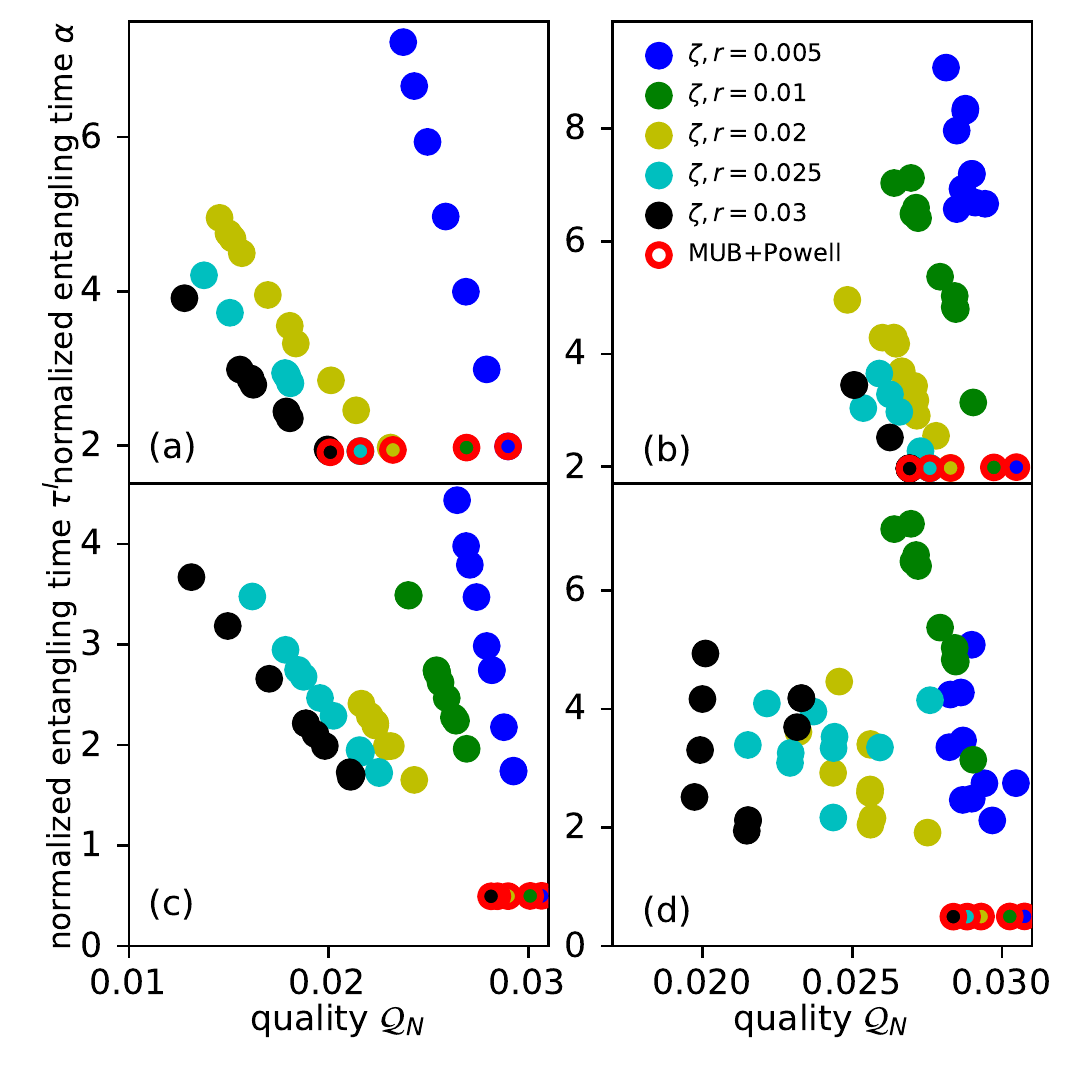}
    \caption{Selected high-performing quorums of projection operators, discovered by optimizing via  simulated annealing for exploration and Powell's local search method for exploitation for the Heisenberg (a,b) and the Ising interaction (c,d) with noise described by the depolarizing channel (a,c) or over- and under-rotation (b,d). The solutions are colored differently, based on the noise level used for each noise model and interaction, where $\zeta$ and $r$ are the noise parameters for the depolarizing channel and over-and-underrotation respectively. 
    The points with the red circles represent the results from using Powell's method with the standard set of as a starting points being as good as or better than the best results from the exploratory analysis. Their colors also represent the noise level of the corresponding noise parameter.}
    \label{fig:exploratoryglobal}
\end{figure}

The results of the exploratory analysis for the Ising interaction with over- and under-rotation errors reveal that there are quorums with the same $\mathcal{Q_N}$ but different $\tau^I$, see Fig.~\ref{fig:exploratoryglobal} (d).
This is a consequence of the Ising interaction leaving specific product bases unchanged.
For example, the standard basis $\{|00\rangle,|01\rangle,|10\rangle,|11\rangle\}$ is not affected by the Ising-based two-qubit operator in Eq.~(\ref{eq:Ising2qgate}) for $\beta_{jx}=\beta_{jy}=0$ and arbitrary $\beta_{jz}$.

\subsection{Results from using the standard set of MUBs  as a starting point of a local search}
\label{sec:resultsMUBstart}
\begin{figure*}
    \centering
    \includegraphics[width=0.95\textwidth]{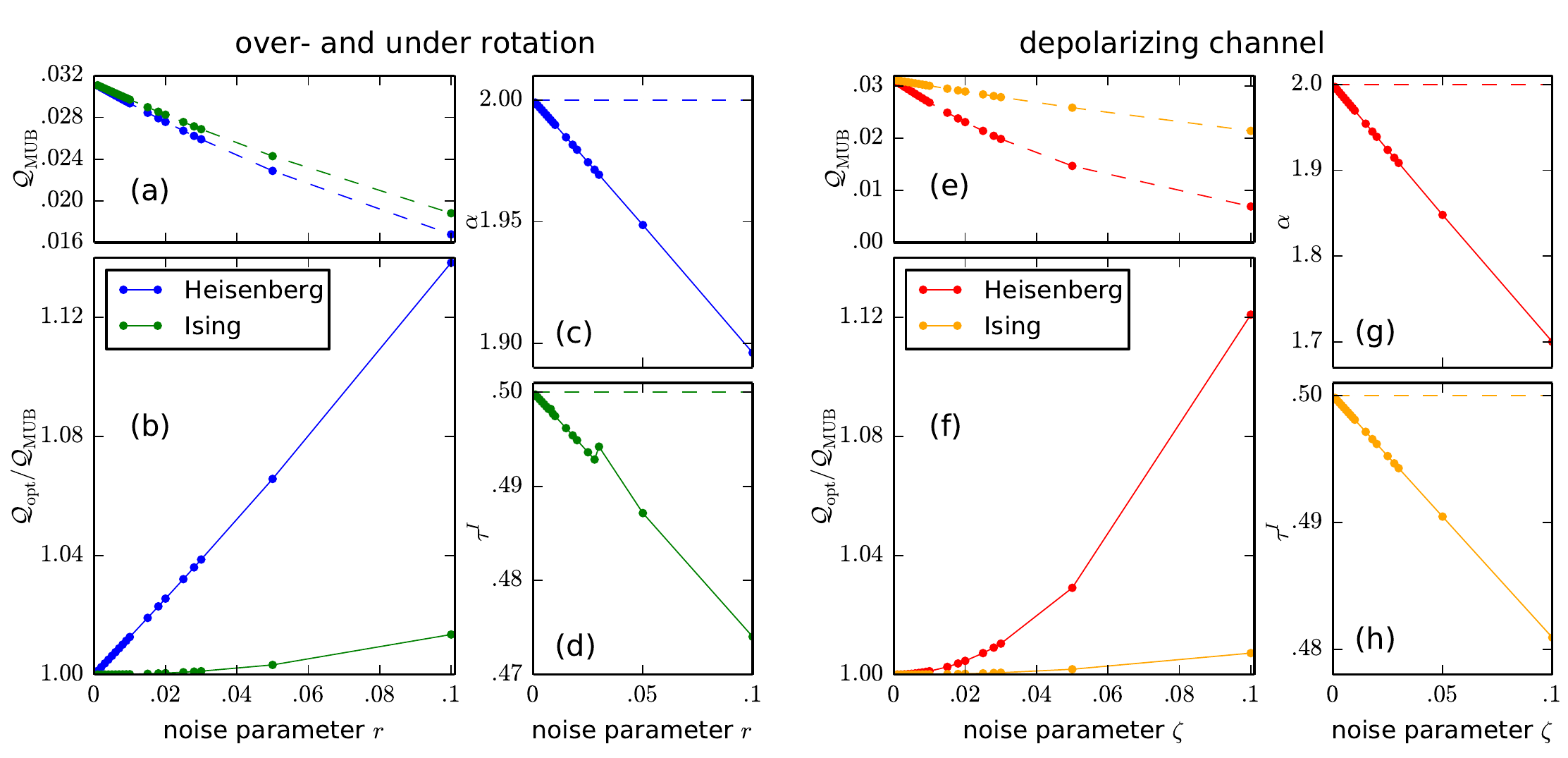}
    \caption{Results for both noise channels, the over- and under-rotation (a-d) and the depolarizing channel (e-h) and for both interactions, the Heisenberg interaction (blue, red) and the Ising interaction (green, orange):
    We present the quality measure $\mathcal{Q}_{\rm MUB}$ for the quorum formed by a complete set of MUBs with lowest possible entangling time (a,e);
    the ratio of the quality measures for the numerically optimized quorum, $\mathcal{Q}_{\rm opt}$ and $\mathcal{Q}_{\rm MUB}$ (b,f);
    as well as the normalized entangling times $\alpha$ (for Heisenberg interaction) (c,d) and $\tau^I$ for the Ising interaction (g,h).
    All these quantities are given as a function of the noise parameters $r$ (a-d) and $\zeta$ (e-h).
    Note that the solid and dashed lines connecting the data points are guidance for the eyes while the dashed lines in (c,d,g,h) represent the (smallest possible) normalized entangling time for the MUBs.
    The increase in $\tau^I$ from $r=0.28$ to $r=0.3$ might be due to the fact that $\mathcal{Q}_N$ can be invariant under switching on the Ising-interaction for certain states, see main text}
    \label{fig:qualityentanglingtime}
\end{figure*}

In Fig.~\ref{fig:qualityentanglingtime}, we present $\mathcal Q_N$ for the MUB quorums in dependence of the noise parameters $r$ and $\zeta$, the improvement by the numerical optimization, and the changes in the normalized overall times the entangling gates are switched on during the QST procedure, $\alpha$ and $\tau^I$. We use here MUB quorums with the presumably minimal amount of time the entangling gates are switched on, see Secs.~\ref{sec:AnalysisHIdep} and \ref{sec:AnalysisZZdep} for details.
As expected, the values of $\mathcal Q_N$ monotonously decrease with increasing $r$ and $\zeta$.
The improvement by the numerical optimization compared to the MUBs increases with an increasing noise level which is due to the fact that the MUBs are ideal at zero noise. In parallel, the normalized entangling times of the numerically optimized quorums decrease.

\subsection{Fidelity of reconstruction}
\label{sec:resultsfidelityofreconstruction}
\begin{figure}
    \centering
    \includegraphics[width=\columnwidth]{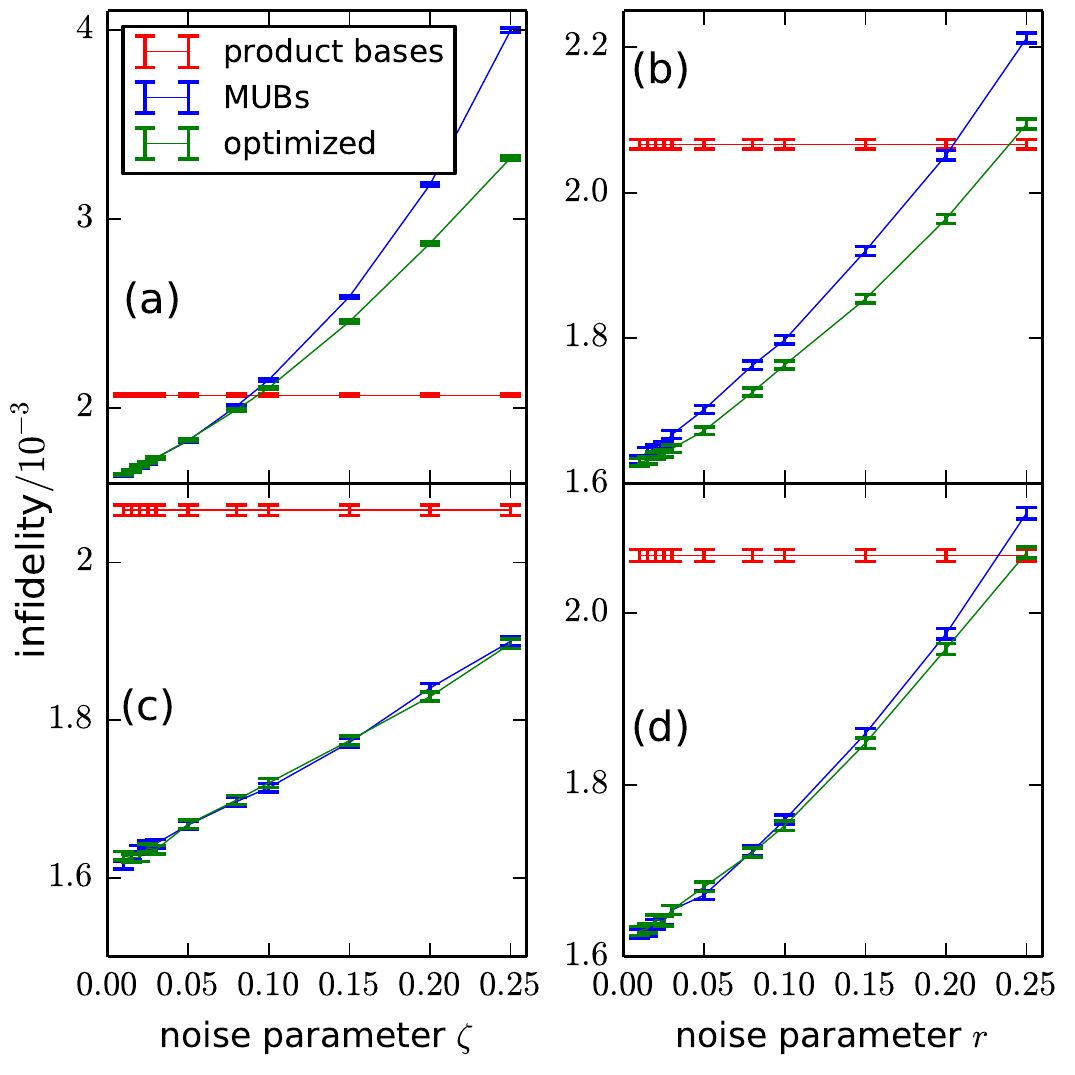}
    \caption{Infidelity of reconstruction in dependence of the noise parameters $\zeta$ and $r$ for the Heisenberg interaction (a,b) and the Ising interaction (c,d) under depolarizing noise (a,c) and over- and under-rotation (b,d). The QST measurement sets were 9 separable bases (red) which don't depend on the noise as only entngling gates are effected, MUBs (blue), and numerically optimized quorums (green). The error bars indicate the standard deviation of the mean after averaging over $10^5$ random density matrices. The total number of measurements runs is $N_{\rm tot}=5\times9\times512$.}
    \label{fig:infidjointplot}
\end{figure}

 We simulate QST by randomly generating density matrices (in the same way as described in Sec.~\ref{sec:effect_of_noise}) and ``measurement outcomes" for a number $N_{\rm rep} = 5\times 9\times512$ of runs of measurements. We use the maximum-likelihood method for reconstruction for $10^5$ random density matrices and present the averaged results in Fig.~\ref{fig:infidjointplot}. The noise levels considered are up to $\zeta=0.25$ and $r=0.25$.
While improvement by the optimization procedure compared to the MUBs increases with increasing $\zeta$ and $r$, the optimization might become imprecise for larger values of $\zeta$ and $r$ as the linearization in Eqs.~(\ref{eq:averaging-4d-expval}) are no longer a good approximation at those values.
However, we are mainly interested in systems with high-purity gates, and, therefore, do not consider higher-order quality measures.

We find that for the depolarizing channel, with Heisenberg interaction up to $\zeta=0.08$, there is a benefit of using entangled-state bases compared to product-state bases.
While for the Ising interaction values of $\zeta$ where QST with nine separable bases performs better than QST with MUBs or numerically optimized quorums, have not been considered, we predict the limiting value to be four times as high as for the Heisenberg interaction,
because the results for Ising and Heisenberg interaction coincide if $\zeta$ for the Ising interaction is four times as high as for the Heisenberg interaction.

For the over- and under-rotation noise model, the fidelity of reconstruction using entangling gates is better than using nine product-state bases for a level of noise around $r=0.2$ for both, Ising and Heisenberg interactions.  
In order to evaluate the benefits of using the entangling gates for performing efficient QST, we need to evaluate how the noise levels in the two models compare to the noise in real state-of-the-art devices and whether for these real-life noise values using entangling gates leads to better performance in comparison to using nine product-state bases.
Therefore, we calculate the relationship between average gate fidelity and the noise parameters of the two models. Using the relationship described in Sec.~\ref{sec:non-degeneratemeasurements}, the average gate fidelity of the CNOT gate which corresponds to the thresholds for the depolarized channel is 0.83. For CNOT gates with this fidelity or higher, we find that the use of entangling gates for measurement is beneficial.
The gate fidelity which corresponds to the threshold $r=0.2$ for the over- and under-rotation channel is $0.85$ or $0.89$ for the Heisenberg and for the Ising interaction, respectively.
For experimentally achievable noise levels \cite{Huang}, there is therefore  already a benefit of using entangling gates when determining optimal QST measurement schemes.

\subsection{Analytical expression for Heisenberg interaction, depolarizing channel}
\label{sec:AnalysisHIdep}
A set of MUBs is obtained by using the unitaries
\begin{equation}
\label{eq:HeisenbergMUB}
    \begin{split}
        U_1= & \mathbbm{1},\\
        U_2 = & U^{q1}(\pi/4,0,0)U^{q2}(\pi/4,0,0),\\
        U_3 = & U^{q1}(\pi/4,0,\pi/2)U^{q2}(\pi/4,0,\pi/2),\\
        U_4 = & U^{q1}(0, \textstyle\frac{\pi}{4}, 0) U^{q2}(\frac{-\pi}{2}, 0, \frac{\pi}{4}) U^{tq}(\frac{1}{2}, 0, \frac{1}{2})\\
              & \times U^{q2}(\frac{\pi}{4}, \pi, -\pi),\\
        U_5 = & U^{q1}(\textstyle\frac{\pi}{4}, \frac{\pi}{4}, \frac{\pi}{4}) U^{q2}(0, \frac{\pi}{4}, 0) U^{tq}(\frac{1}{2}, 0, \frac{1}{2}).
    \end{split}
\end{equation}
Entangling gates are used only for $U_4$ and $U_5$ with the parameters $\alpha_{41}=\alpha_{43}=\alpha_{51}=\alpha_{53}=1/2$ and $\alpha_{42}=\alpha_{52}=0$.
From our numerical results, we observe that for the optimized quorum only the values of $\alpha_{41}$, $\alpha_{43}$, $\alpha_{51}$, and $\alpha_{53}$ change in dependence of $\zeta$.
Therefore, we try to reproduce the numerical result analytically by fixing all the parameters of the single-qubit gates included in a quorum according to Eq.~(\ref{eq:HeisenbergMUB}) and $\alpha_{42}=\alpha_{52}=0$ but treat $\alpha_{41}$, $\alpha_{43}$, $\alpha_{51}$, and $\alpha_{53}$ as independent parameters.
Then the quality measure reads
\begin{equation}
\begin{split}
    \mathcal{Q}_N = & \frac{\sin(\alpha_{41}\pi)\sin(\alpha_{43}\pi)}{32}\\
    &\times \cos^4\left(\frac{(\alpha_{51}{-}\alpha_{53})\pi}{2}\right)\sin^2\left(\frac{(\alpha_{51}{+}\alpha_{53})\pi}{2}\right)\\
    & \times\exp[-\zeta s(\alpha_{41}+\alpha_{43}+\alpha_{51}+\alpha_{53})].
    \end{split}
\end{equation}
Although the expression is not symmetric under the exchange of the values of $\alpha_{4k}$ and $\alpha_{5k}$ ($k=1,3$), the maximum is reached at a point where all parameters coincide,
\begin{equation}
    \alpha^{\rm max}_{41} = \alpha^{\rm max}_{43} = \alpha^{\rm max}_{51} = \alpha^{\rm max}_{53} = \frac{1}{\pi}\arctan\left(\frac{1}{\zeta s}\right).
\end{equation}
With this expression, we reproduce the numerical results.

\subsection{Analytical expression for Ising interaction, depolarizing channel}
\label{sec:AnalysisZZdep}
The same set of MUBs obtained above for the Heisenberg interaction can be realized using the Ising interaction, by representing the two-qubit gates within the sequence for $U_4$ and $U_5$ in Eq.~(\ref{eq:HeisenbergMUB}) with the parameters $\beta_{4y}=\beta_{5y}=\pi/4$ and $\beta_{4x}=\beta_{4z}=\beta_{5x}=\beta_{5z}=0$.
Again, we observe within the numerics that with increasing values of $\zeta$ all parameters but $\beta_{4y}$ and $\beta_{5y}$ remain the same.
Thus, we consider the quorum as described above treating $\beta_{4y}$ and $\beta_{5y}$  as free parameters.
We obtain for the quality measure
\begin{equation}
    \mathcal{Q}_N=\frac{\sin^2(2\beta_{4y})\sin^2(2\beta_{5y})}{32}e^{-\zeta s(\beta_{4y}+\beta_{5y})}
\end{equation}
and find the maximum at 
\begin{equation}
    \beta^{\rm max}_{4y}=\beta^{\rm max}_{5y}=\frac{1}{2}\arctan\left(\frac{4}{\zeta s}\right)
\end{equation}
which indeed yields the same $\mathcal{Q}_N$ as the numerically optimized solution for each $\zeta$.

\subsection{Performance of QST with  Noisy Gates on Real  Quantum Device}
\begin{figure*}
    \centering
    \includegraphics[width=\columnwidth]{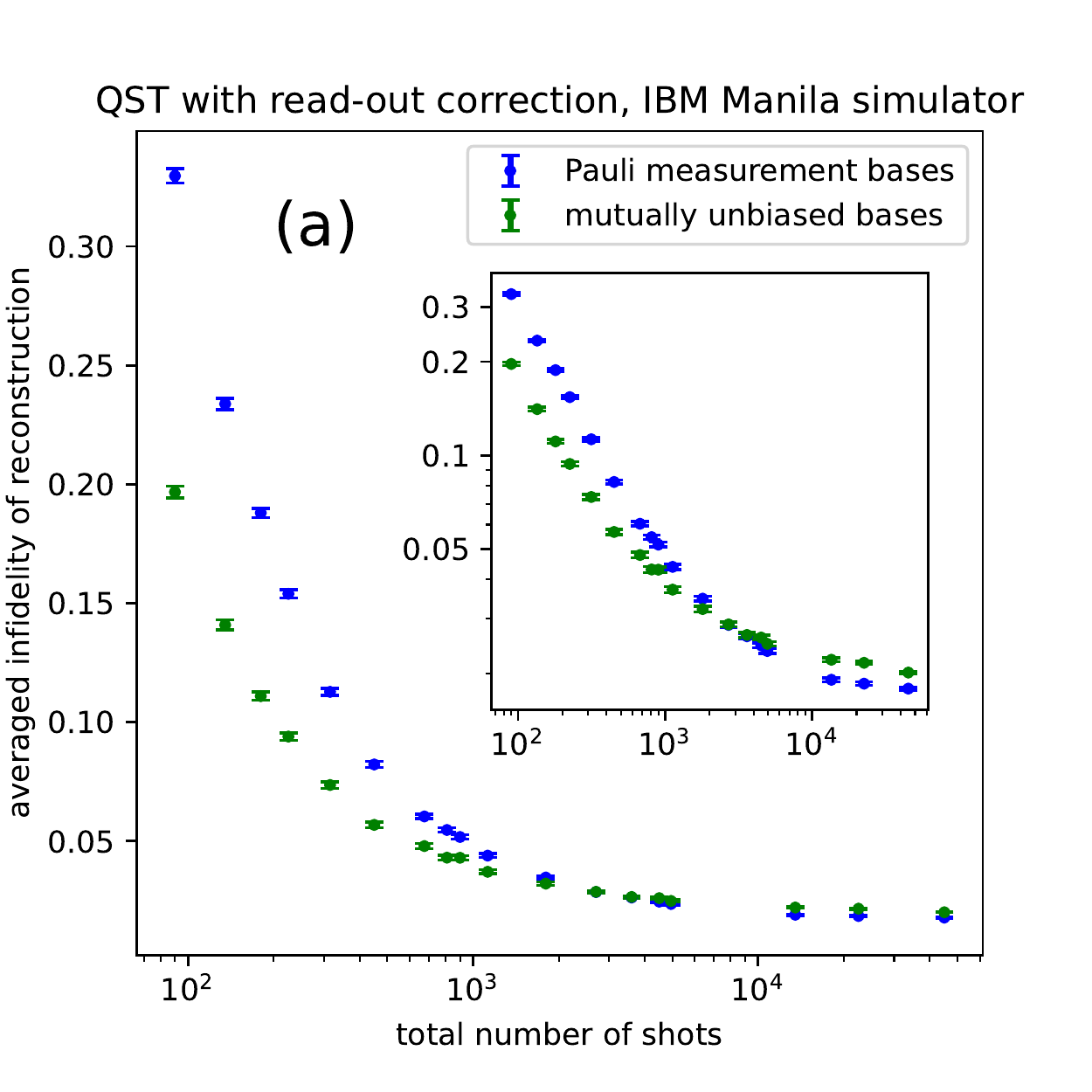}\includegraphics[width=\columnwidth]{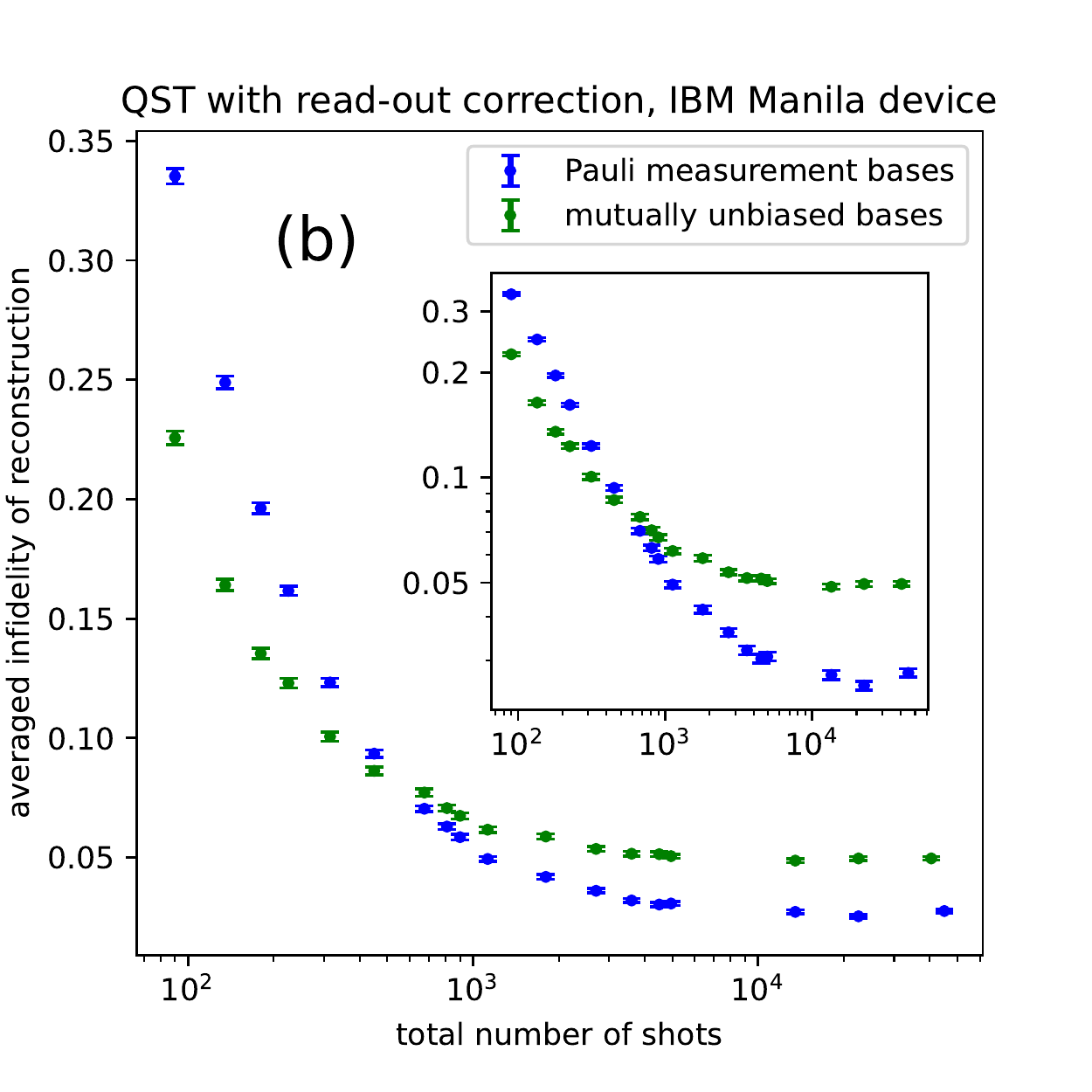} 
    \caption{Averaged infidelity of reconstruction for 660 pure initial states for QST with nine Pauli product bases (blue) and five MUBs (green) for the IBM Manila simulator  (a) and the IBM Manila quantum processor (b).
    We observe that the MUBs outperform the Pauli bases for a low number of shots, while the Pauli bases perform better for a higher number of shots possibly due to not including any two-qubit gate within the measurement scheme.  Insets: Log-log plots of the same averaged infidelities.}
    \label{fig:Manila}
\end{figure*}

Given that the  noise models considered above require stringent limiting assumptions on the behavior of the noise rather than using noise characteristics derived from a real quantum computer, we decided to examine our approach using an actual noisy quantum device.
However, running our optimized measurement bases on the publicly available quantum devices is challenging, due to the fact that a $\mathrm{SWAP}^\alpha$ gate with $\alpha$ being a parameter is not available as a native gate, CNOT beign the only native two-qubit gate.
Reproducing $\mathrm{SWAP}^\alpha$ would require a gate sequence with a fixed number of CNOTs, i.e., $\alpha$ would not determine the time of the two-qubit interaction being switched on.
Therefore, for the two-qubit noisy system, which was the focus of investigation in this paper, we ran comparative QST using -- as measurement bases -- only a full set of MUBs and Pauli product bases.
Using Pauli bases is currently the state of the art.
In order to evaluate how the two approaches using different measurement bases are affected specifically by the noisy gates, we corrected the readout error in both cases before the QST was performed.  
We ran QST using both measurement bases on a classical simulation of the IBM Manila device, as well as on the actual IBM Manila quantum device.
It is important to note that the simulation, which uses averaged data, may be more representative than the results run on the actual device, which may differ from calibration to calibration. 

We ran both QST approaches for different total numbers of allowed repetitions, $N_{\rm tot}$, i.e. the total number of \emph{shots} using the qiskit terminology, and evaluated the results.
The results, averaged over 660 randomly selected pure initial states, are shown in Fig.~\ref{fig:Manila} (a) for the simulated Manila quantum device, and in Fig.~\ref{fig:Manila} (b) for the actual quantum device.
For the simulated data, for a fidelity of reconstruction of 97\%, the MUBs outperform Pauli bases as measurement bases and require a lower number of repetitions $N_{\rm tot}$.
The crossover happens at a total number of shots of around $N_{\rm tot}=2\times10^3$.
It is important to note that for further increasing $N_{\rm tot}$ the fidelity saturates and the gain from performing extra shots is very small, even if the Pauli bases are more advantageous.  In the case of the quantum processor, the MUBs outperform Pauli bases up to a fidelity of reconstruction close to 92\%, achieved for around $N_{\rm tot}=500$; for larger $N_{\rm tot}$ the Pauli bases are the more advantageous approach to use.
Note that the averaged infidelity of reconstruction saturates for large numbers of measurement repetitions due to systematic errors.
Systematic errors in the CNOT gates are most likely to yield the QST with MUBs having a higher saturation value for the infidelity.
This explains the crossover behavior: for lower values of $N_{\rm tot}$ where the infidelity is not dominated by systematic errors, our analysis that MUBs outperform QST with Pauli bases is confirmed; when the systematic errors dominates, Pauli bases are better as they do not suffer from imperfect CNOT gates included in measurement scheme.
It should be mentioned that these are results achieved using one calibration of the quantum device, and not an averaged performance over several calibrations, and thus we consider the results of the simulator as more representative. 

Clearly, for scenarios where the total number of shots is limited, such as when one needs to test initialization of a  quantum device and thus a large number of random initial density matrices, the  QST with MUBs is the more beneficial approach. 
An additional advantage of the QST with MUBs as measurement bases comes from the fact that fewer different circuits are involved, and currently the loading of new circuits is one of the main bottlenecks of using the IBM quantum devices.
The advantage of having a smaller number of different measurements (different circuits) is increasing for increasing number of qubits, scaling as a ratio of $(2/3)^n$ where $n$ is the number of qubits. Therefore, there is an exponential advantage of MUBs over Pauli product bases regarding the number of different measurements.

\subsection{Discussion of Results}
We performed an extensive analysis of the search space of quorums for QST, using global and local parallel explorations. However, the best solutions we were able to discover are the ones that are obtained by a local search method using a full set of MUBs as a starting point.
Additionally, these optimized solutions differ from the standard set of MUBs only by different entangling times. 
Using the derived quality measures, for realistic noise levels, we find that the optimized solutions are better than the MUBs.  For realistic depolarizing and over- and under-rotation noise models using the Heisenberg interaction, there is a small improvement in the fidelity of the reconstruction results for the optimized quorum compared to the standard set of MUBs. For the Ising interaction (depolarized and over- and under-rotation), there is no improvement over using the MUBs.

The state-of-the-art approach for efficient QST under noise is to use nine separable bases. Our results demonstrate that this is unnecessary. Namely, the standard set of  MUBs performs significantly better in state-of-the-art existing systems.  

\section{Methods}
\label{sec:Methods}
\subsection{Parametrization of non-degenerate measurements}
\label{sec:non-degeneratemeasurements}
A quantum gate according to Eq.~(\ref{eq:universal2-q-gate}) is given by 15 real parameters.
For a QST measurement set we need five different quantum gates.
Therefore, we have overall 75 parameters.
We consider noise caused by the entangling gates while we assume that the single-qubit gates are error-free and that they can be performed instantly.

\subsubsection{QST for two qubits with noisy entangling gates, over- and under-rotation}
For the noise of the entangling gate, we include over- and under-rotation where the parameters $\alpha_{j,m}$ follow a Gaussian distribution  while the entangling gate is switched on, focusing for now on the Heisenberg interaction.
The noise-affected operation can be denoted as the desired unitary operation followed by the linear positive map in the eigenbasis of $U^{tq}$
\begin{equation}
    \rho \mapsto
\left(\begin{array}{cccc}
\rho_{00} & \gamma_1\rho_{01} & \gamma_2\rho_{02} & \gamma_3\rho_{03}\\
\gamma_1\rho_{10} & \rho_{11} & \gamma_1\gamma_2\rho_{12} & \gamma_1\gamma_3\rho_{13}\\
\gamma_2\rho_{20} & \gamma_1\gamma_2\rho_{21} & \rho_{22} & \gamma_2\gamma_3\rho_{23}\\
\gamma_3\rho_{30} & \gamma_1\gamma_3\rho_{31} & \gamma_2\gamma_3\rho_{32} & \rho_{33}
                      \end{array}\right)
\end{equation}
with $\gamma_j=e^{-r\alpha_j\pi}$.
We can express this map by eight Kraus operators
\begin{equation}
    \begin{split}
    M_{mkl} = &\sqrt{\frac{(1+(-1)^m\gamma_1)(1+(-1)^k\gamma_2)(1+(-1)^l\gamma_3)}{8}}\\
              & \times \operatorname{diag}(1,(-1)^m,(-1)^k,(-1)^l)
    \end{split}
\end{equation}
\\
for $m,k,l\in\{0,1\}$.
From the Kraus operators, the average gate fidelity can be directly computed \cite{Pedersenetal}
\begin{equation}
    F^{\rm Heisenberg}_{\rm ou} = \frac{4\!+\!\gamma_1\!+\!\gamma_2\!+\!\gamma_3\!+\!\gamma_1\gamma_2\!+\!\gamma_1\gamma_3\!+\!\gamma_2\gamma_3}{10}.
\end{equation}
For a CNOT gate realized by $\alpha_1=\alpha_3=1/2$ and $\alpha_2=0$, we obtain
\begin{equation}
    F^{\rm Heisenberg}_{\rm ou}({\rm CNOT}) = \frac{1}{2} + \frac{2}{5}e^{-r\pi/2} + \frac{1}{10}e^{-r\pi}.
\end{equation}
This allows us to compare the noise parameter $r$ from our model to  average gate fidelities of existing implementations of qubit systems.

For the Ising interaction, we obtain, in the Bell basis, a modified map for the effect of the noise
\begin{equation}
\label{eq:Ising-map}
    \rho\mapsto
    \left(\begin{array}{cccc}
\rho_{00}          & \rho_{01} \gamma_y\gamma_z & \rho_{02}\gamma_x\gamma_y & \rho_{03}\gamma_x\gamma_z \\
\rho_{10}\gamma_y\gamma_z  & \rho_{11}          & \rho_{12}\gamma_x\gamma_z & \rho_{13}\gamma_x\gamma_y \\
\rho_{20}\gamma_x\gamma_y  & \rho_{21}\gamma_x\gamma_z  & \rho_{22}         & \rho_{23}\gamma_y\gamma_z  \\
\rho_{30}\gamma_x\gamma_z  & \rho_{31}\gamma_x\gamma_y  & \rho_{32}\gamma_y\gamma_z & \rho_{33}   
    \end{array}
    \right)
\end{equation}
with
$\gamma_j=e^{-2r|\beta_j|}$.
This map is represented by the four Kraus operators
\begin{equation}
        \begin{split}
M_{kl} = &\sqrt{\frac{1{+}(-1)^k\gamma_y\gamma_z{+}(-1)^l\gamma_x\gamma_y+(-1)^{k+l}\gamma_x\gamma_z}{4}}\\
         & \operatorname{diag}(1,(-1)^k,(-1)^l,(-1)^{k+l}),
    \end{split}
\end{equation}
for $k,l\in\{0,1\}$.
The averaged gate fidelity is given by
\begin{equation}
    F_{\rm ou}^{\rm Ising} = \frac{2+\gamma_x\gamma_y+\gamma_y\gamma_z+\gamma_x\gamma_z}{5}.
\end{equation}
For the CNOT gate where $\beta_z=\pi/4$, we obtain
\begin{equation}
     F_{\rm ou}^{\rm Ising} ({\rm CNOT}) = \frac{3}{5} +\frac{2}{5}e^{-r\pi/2}.
\end{equation}

Note that the noise does not affect all states of a measurement basis in the same way as the gate does not entangle each product input state.
This means our considerations from Sec.~\ref{sec:effect_of_noise} need to be adjusted.
While a measurement $j$ in the presence of noise is still described by a POVM, $\{F_{j1},F_{j2},F_{j3},F_{j4}\}$, Eq.~(\ref{eq:proj-noise}) does not hold anymore.
However, we can find projectors $P_{jk}'$ with $k=1,2,3,4$ such that
$F_{jk} = q_{jk}(P'_{jk}-l_{jk}\mathbbm{1}/d) + l_{jk}\mathbbm{1}/d$,
where $q_{jk}$ now explicitly depends on $k$, and can be extracted from the $F_{jk}$ by
\begin{equation}
q_{jk} = \sqrt{\frac{4}{3}\tr\left[\left(F_{jk}-\frac{{\mathbbm{1}}}{4} \right)^2\right]}.
\end{equation}
Note that the rank-1 projectors $P_{jk}'$ do not necessarily form an orthogonal basis.
However, when we select three from each of the five measurements, the volume $\mathcal{Q}$ spanned by those 15 projectors does not depend on the selection of the three out of four basis states.
The noise-affected quality measure is then given by
\begin{equation}
    \mathcal{Q}_N = \mathcal{Q}\prod_{j=1}^5\prod_{k=1}^4 q_{jk}^{1.195/2},
\end{equation}
obtaining the exponent from Eq.~(\ref{eq:averaging-4d-expval}).

\subsubsection{QST for two qubits with noisy entangling gates - depolarizing channel}
Depolarization leads to the exponential decay of all of the components of the resulting density matrices which are not proportional to the identity matrix,
\begin{equation}
    \rho \mapsto q_j(\rho-\mathbbm{1}/4) + (1-q_j)\mathbbm{1}/4.
\end{equation}
This map is expressed by the Kraus operators
\begin{equation}
    M_{kl} = \frac{\sqrt{1-q_j}}{4}\sigma_k\otimes\sigma_l
\end{equation}
for $k,l\in\{0,x,y,z\}$ but $(k,l)\ne(0,0)$ and
\begin{equation}
       M_{00} =  \frac{\sqrt{15q_j+1}}{4}\mathbbm{1}.
\end{equation}
Using again the formalism from \cite{Pedersenetal} this yields an average gate fidelity of
\begin{equation}
    F_{d} = \frac{1+3q_j}{4}.
\end{equation}
The probability to leave the state unchanged by the depolarization is given by
\begin{equation}
q_j=\exp\left(-\zeta\pi\sum_{k=1}^3\alpha_{jk}\right) =c\exp\left(-\zeta\sum_{k=x,y,z}\beta_{jk}\right)
\end{equation}
for the Heisenberg and the Ising interaction respectively, $\zeta$ is a measure for how strongly the noise affects the quantum system.
For the CNOT gate we then obtain the average gate fidelities
\begin{equation}
    F_d^{\rm Heisenberg} ({\rm CNOT}) = \frac{1}{4} + \frac{3}{4} e^{-\zeta\pi}
    \label{eq:averagegateFidelityHIdep}
\end{equation}
and
\begin{equation}
    F_d^{\rm Ising} ({\rm CNOT}) = \frac{1}{4} + \frac{3}{4} e^{-\zeta\pi/4}.
    \label{eq:averagegateFidelityZZdep}
\end{equation}
In order to obtain a rough idea of what range the noise level might be in a realistic scenario, we make the strong assumption that the depolarizing channel would describe the noise in a system correctly.
Then,  we could extract $\zeta$ from experimental estimations of the average gate fidelity, e.g. the value of $F_{\rm Huang}=0.98$  as found for a CNOT-equivalent gate in \cite{Huang},
$\zeta_{\rm Huang} = 0.034$.
Note that the over-and-under-rotation picture as we consider it here for the Ising interaction cannot directly be related to the results in \cite{Huang} as the two-qubit interaction there comes along with a single-qubit rotations.

The considerations from Sec.~\ref{sec:effect_of_noise} can be applied in a straightforward manner for the depolarizing channel.

\subsubsection{Comparison to entanglement-free QST}
We compare a QST quorum including entangling gates to QST without entanglement.
However, there is no informationally complete set of five measurement operators whose eigenbases include only product states.
A standard procedure with separable basis states only is a set of nine measurements given by all combinations of measuring in the Pauli $x$, $y$, and $z$ bases for the first and the second qubit \cite{Steffenetal2006,Zajacetal2018,Watsonetal2018}.
We simulate quantum measurements and compute the fidelity of reconstruction for the nine measurements without entanglement and the entanglement-including quorums.

\subsection{Numerics: Methods}
\label{sec:numerics}
\subsubsection{Exploratory Analysis}
The problem of finding the optimal quorum of projection operators under noise is a non-convex continuous optimization problem with the derivatives of the function not easily obtained, and with multiple local maxima.
From our previous work \cite{VioletaNiklas,VioletaGuidoNiklas}, we know that using a local optimizer (Powell's method) with well-chosen starting points performs very well.
Here, we use the Powell's derivative free method started in parallel with multiple sufficiently diverse starting points in order to improve the exploration of the search space. In addition, we used a global optimization approach. Based on the results of the exploratory analysis and the theoretical considerations in \ref{sec:standardMUBstartingpoint}, to discover the optimized quorums for each noise level, we used a local search approach, with starting point the standard set of MUBs.   

\subsubsection{Local search: Powell's method}
Initially, Powell's method for local search with $500$ well chosen starting points was used. 
The points were chosen at random, but with the requirement to meet a diversity threshold, where the diversity was evaluated using the  the angles formed by the traceless parts of the projection operators. For detail see below. The diversity measure was based on the Jaccard distance. The distance between two quorums was considered to be the normalized minimal Jaccard distance that each projector from a quorum $Q_1$ forms with the projectors from the quorum $Q_2$ based on the angles formed by the traceless parts of the projection operators. The chosen diversity measure threshold used for each of the noise models and interactions is different, based on the distribution of distances between two randomly chosen quorums (the threshold was chosen to be the mean - one standard deviation). 

\subsubsection{Diversity measure for the local search}
In the absence of noise, the quality of a quorum is uniquely determined by the pairwise dot products of the traceless part $Q_{jk}=P_{jk}-\mathbbm{1}/4$ of the projection operator $P_{jk}$ projecting on the eigenstate $k$ of the measurement operator $j$, treating the $Q_{jk}$ as vectors in a vector space with the dot product $\tr(Q_{jk}Q_{j'k'})$. Thus, a diversity measure based on these dot products of two quorums is meaningful. Here, we are interested in small levels of noise of up to $\zeta=0.03$ and $r=0.03$, for which we know that the effect of the noise levels on the quality measure $\mathcal{Q}_N$ is small, and distance measure based only on dot products of the $Q_{jk}$ is still valid.  

Then, each quorum is a set of sets, which are the dot products that each $Q_{jk}$ forms with the other $Q_{j'k'}$ in the quorum. As a distance measure between two quorums we use the Jaccard distance \cite{jaccard}. 

\subsubsection{Global Optimization: Simulated Annealing}
As a part of the exploratory analysis, we performed global optimization using Simulated Annealing with appropriately selected parameters, in combination with a local search approach (Powell's method). This combination is commonly used with success when solving an optimization problem with a complicated landscape: the global optimizer is used for exploration of the search space and locating promising areas, while the local optimizer is then used for exploitation, i.e. refinement and reaching the closest locally optimal point. 

\subsubsection{Using the standard set of MUBs as a starting point of the local search}
\label{sec:standardMUBstartingpoint}
For zero noise, MUBs are known to be the ideal choice for an QST quorum and the noise penalizes switching on entangling gates. Thus one can expect for small noise the optimal solutions to be close to the set of MUBs with minimal entangling times.
Indeed, many of the best solutions obtained during the exploratory analysis have entanglement, which corresponds to two bases with entangled states. There is a known set of MUBs constructed via the approach presented in \cite{Klappenecker2003}, known as the standard set of MUBs, where two bases are with maximally entangled states. 
Its parametrization using the Heisenberg interactions is given in Eqs.~(\ref{eq:HeisenbergMUB}).

Considering the above, we use the standard set of MUBs as starting points for Powell's method to find the best optimized solutions for low noise level.

\subsection{Running QST on IBM Manila}
We used the interface provided by qiskit to execute QST on the IBM Manila
quantum processor as well as the corresponding simulator.
Using Pauli bases is the default and readily implemented for the state tomography function in qiskit.
We extended the libraries of qiskit to also perform QST based on MUBs in order to compare the two approaches. 
Using the class ``CompleteMeasFitter" \cite{qiskit1}, we corrected the readout error in both cases prior to reconstruction of the QST being performed.  
660 initial states were randomly selected as pure two-qubit states by first normalizing random Gaussian states in $\mathbbm{R}^4$ and then adding three relative phases randomly chosen from a uniform distribution on the interval $[0,2\pi)$.

\section{Conclusions and Overview}
\label{sec:conclusions}
To summarize, we investigated the optimal QST measurement schemes under the influence of noise.
We extended Wootters and Fields' \cite{WoottersFields89} quality measure for a QST measurement quorum to the case of noise-affected measurements and optimized QST measurement sets for a single- and two-qubit system under noise.
For a single qubit, we considered noise which increases with the polar angle of the Bloch sphere and perfect azimuthal rotations.
For two qubits, we limited the discussion to perfect single-qubit gates and noisy two-qubit gates, generated either by Heisenberg interaction or by Ising interaction.
We  solved the problem of finding an optimal quorum for quantum state tomography under these noise models by using an extensive number of well-suited numerical techniques. 

For two qubits, the results depend on the interaction and on the noise model. For practically relevant noise levels, a minor improvement over using MUBs is present for the Heisenberg over- and under-rotation and depolarized noise models. Apart from this, the set of MUBs performs sufficiently well as an quorum for QST for realistic noise levels. 

In some cases, we extracted analytical expressions for the optimized quorum from the numerical results, namely for the single-qubit case and for two qubits with a depolarizing channel. In the two-qubit case, only the entangling gate times change as the noise level is varied.

Importantly, for simulated QST based on the noise models described above we find an improvement of QST with MUBs and numerically optimized QST measurement sets compared to QST with separable bases.
While we did not include state preparation and measurement (SPAM) errors, their influence can be mitigated \cite{Palmerietal2020}.

To confirm our findings and alleviate potential limitations of our noise models, we compared the performance of QST using entangling gates with QST using nine separable bases on a real quantum device.  We investigated for which scenarios the use of entangled gates is advantageous in comparison with the use of the nine product bases as measurement bases.

Naturally, future research would consider 
 models with noisy two-qubit and single-qubit gates, using system-specific parameters.

\section*{Data and code availability}
The code and generated data needed to reproduce the results of this work are publicly available \cite{SupplMat}.

\section*{Acknowldgements}
This work was partially supported by the Zukunftskolleg (University of Konstanz) and the Bulgarian National Science Fund under the contract No KP-06-M 32/1.

\section*{Author contributions}
All authors jointly developed the idea of optimizing QST measurement sets for systems under noise and found the modified quality measure.
NR formulated and solved the optimization problem for a single qubit.
GB suggested the parametrization based on the decomposition of quantum gates. Based on this approach, NR formulated the explicit optimization problem.
VNI-R implemented and applied all approaches described in the Methods section to solve this problem.
NR wrote the code for the obtaining the average infidelity of reconstruction by simulated quantum measurements.
VNI-R and NR extended the qiskit libraries to include QST with MUBs. 
All authors participated in the discussion of the models and of the results as well as in writing the manuscript.


\begin {thebibliography}{45}

\bibitem{Shor}P.~W.~Shor,
Polynomial-time algorithms for prime factorization and discrete logarithms on a quantum computer.
\href{https://doi.org/10.1137/S0097539795293172}{SIAM J.~Sci.~Statist.~Comput.~\textbf{26}, 1484} (1997).

\bibitem{QuantumSimulationsRMP}I.~M.~Georgescu, S.~Ashhab, and F.~Nori,
Quantum simulation.
\href{https://doi.org/10.1103/RevModPhys.86.153}{Rev.\ Mod.\ Phys.\ \textbf{86}, 153} (2014).

\bibitem{ionsreview}C. D. Bruzewicz, J. Chiaverini, R. McConnell,  and J. M. Sage,
Trapped-ion quantum computing: progress and challenges.
\href{https://doi.org/10.1063/1.5088164}{Appl.\ Phys.\  Rev.\ \textbf{6}, 021314} (2019).

\bibitem{Aruteetal2019}F.\ Arute, K.\ Arya, R.\ Babbush, D.\ Bacon, J.\ C.\ Bardin, R.\ Barends, R.\ Biswas, S.\ Boixo, F.\ G.\ S.\ L.\ Brandao, D.\ A.\ Buell \textit{et al.},
Quantum supremacy using a programmable superconducting processor.
\href{https://doi.org/10.1038/s41586-019-1666-5}{Nature \textbf{574}, 505} (2019).

\bibitem{supercondqubitsreview}M.~Kjaergaard, M.~E.~Schwartz, J.~Braum\"uller, P.~Krantz, J.~I.-J.~Wang, S.~Gustavsson, and W.~D.~Oliver,
Superconducting qubits: current state of play.
\href{https://doi.org/10.1146/annurev-conmatphys-031119-050605}{Annual Review of Condensed Matter Physics \textbf{11}, 369 }(2020).

\bibitem{LossDiVincenzo}D.~Loss and D.~P.~DiVincenzo,
Quantum computation with quantum dots.
\href{https://doi.org/10.1103/PhysRevA.57.120}{Phys.~Rev.~A \textbf{57}, 120} (1998).

\bibitem{KloeffelLoss2013}C.~Kloeffel and D.~Loss,
Prospects for spin-based quantum computing in quantum dots.
\href{https://doi.org/10.1146/annurev-conmatphys-030212-184248}{Annual Review of Condensed Matter Physics \textbf{4}, 51} (2013).

\bibitem{NV-centers} P. Nizovtsev, S. Ya. Kilin, F. Jelezko, T. Gaebal, I. Popa, A. Gruber, and J. Wrachtrup,
A quantum computer based on NV centers in diamond: optically detected nutations of single electron and nuclear spins.
\href{https://doi.org/10.1134/1.2034610}{Optics and Spectroscopy \textbf{99}, 233} (2005).

\bibitem{QSTbookchapter2004}J.~B.~Altepeter, D.~F.~James, and P.~G.~Kwiat,\href{https://doi.org/10.1007/978-3-540-44481-7_4}{\textit{Qubit Quantum State Tomography.} In: M. Paris, J.~\v{R}eh\'{a}\v{c}ek (eds.), Quantum State Estimation. Lecture Notes in Physics, \textbf{649}. Springer, Berlin, Heidelberg} (2004).

\bibitem{Emersonetal2005}J. Emerson, R. Alicki, and K. \.{Z}yczkowski,
Scalable noise estimation with random unitary operators.
\href{https://doi.org/10.1088/1464-4266/7/10/021}{J. Opt. B: Quantum Semiclass. Opt. \textbf{7}, S347} (2005).

\bibitem{Mavadiaetal2018}S. Mavadia, C. L. Edmunds, C. Hempel, H. Ball, F. Roy, T. M. Stace, and M. J. Biercuk,
Experimental quantum verification in the presence of temporally correlated noise.
\href{https://doi.org/10.1038/s41534-017-0052-0}{npj Quantum Information \textbf{4}, 7} (2018).

\bibitem{Merkeletal2013}S. T. Merkel, J. M. Gambetta, J. A. Smolin, S. Poletto, A. D. C\'{o}rcoles, B. R. Johnson, C. A. Ryan, and M. Steffen,
Self-consistent quantum process tomography.
\href{https://doi.org/10.1103/PhysRevA.87.062119}{Phys.\ Rev.\ A \textbf{87}, 062119} (2013).

\bibitem{Blume-Kohoutetal2013}R. Blume-Kohout, J. K. Gamble, E. Nielsen, J. Mizrahi, J. D. Sterk, and P. Maunz,
Robust, self-consistent, closed-form tomography of quantum logic gates on a trapped ion qubit.
Preprint at \href{https://arxiv.org/abs/1310.4492}{arXiv:1310.4492}.

\bibitem{Nielsenetal2020}E. Nielsen, J. K. Gamble, K. Rudinger, T. Scholten, K. Young, and R. Blume-Kohout,
Gate set tomography.
Preprint at \href{https://arxiv.org/abs/2009.07301}{arXiv:2009.07301}.

\bibitem{Crameretal2010}M.\ Cramer, M. B. Plenio, S. T. Flammia, R. Somma, D. Gross, S. D. Bartlett, O. Landon-Cardinal, D. Poulin, and Y.-K. Liu,
Efficient quantum state tomography.
\href{https://doi.org/10.1038/ncomms1147}{Nature Communications \textbf{1}, 149} (2010).

\bibitem{WoottersFields89}W.~K.~Wootters and B.~D.~Fields,
 Optimal state-de\-ter\-mination by mutually unbiased measurements.
 \href{https://doi.org/10.1016/0003-4916(89)90322-9}{Ann.~Phys.~\textbf{191}, 363} (1989).
 
\bibitem{Rehaceketal2004}J.~\v{R}eh\'a\v{c}ek, B.-G.~Englert, and D.~Kaszlikowski,
 Minimal qubit tomography.
\href{https://doi.org/10.1103/PhysRevA.70.052321}{Phys.\ Rev.\ A \textbf{70}, 052321 } (2004).

\bibitem{Renesetal}J.~M.~Renes, R.~Blume-Kohout, A. J. Scott, and C.~M.~Caves,
Symmetric informationally complete quantum measurements.
\href{https://doi.org/10.1063/1.1737053}{J.~Math.~Phys.~\textbf{45}, 2171} (2004).
 
\bibitem{BodmannHaas2018}B.~G.~Bodmann and J.~I.~Haas,
Maximal orthoplectic fusion frames from mutually unbiased bases and block designs.
\href{https://doi.org/10.1090/proc/13956}{Proc. Amer. Math. Soc.~\textbf{146}, 2601} (2018).
 
\bibitem{VioletaNiklas}V.~N.~Ivanova-Rohling and N.~Rohling,
Optimal choice of state tomography quorum formed by projection operators.
\href{https://doi.org/10.1103/PhysRevA.100.032332}{Phys.~Rev.~A \textbf{100}, 032332} (2019).

\bibitem{VioletaGuidoNiklas}V.~N.~Ivanova-Rohling, G.~Burkard, and N.~Rohling,
Quantum state tomography as a numerical optimization problem,
\href{https://doi.org/10.1088/1367-2630/ac3c0e}{New J.~Phys.~\textbf{23} 123034} (2021).

\bibitem{deBurgh2008} M.~D.~de Burgh, N.~Langford, A.~Doherty, and A.~Gilchrist,
Choice of measurement sets in qubit tomography.
\href{https://doi.org/10.1103/PhysRevA.78.052122}{Phys. Rev. A \textbf{78}, 5} (2008)

\bibitem{Mohammadi2014} M.~Mohammadi and A.~Bra\ifmmode \acute{n}\else \'{n}\fi{}czyk,
Optimization of quantum state tomography in the presence of experimental constraints.
\href{https://doi.org/10.1103/PhysRevA.89.012113}{Phys. Rev. A \textbf{89}, 1} (2014)

\bibitem{Miranowicz2014} A.~Miranowicz, K.~Bartkiewicz, J.~Pe\ifmmode \check{r}\else \v{r}\fi{}ina, M.~Koashi, N.~ Imoto, and F.~Nori,
Optimal two-qubit tomography based on local and global measurements: Maximal robustness against errors as described by condition numbers.
\href{https://doi.org/10.1103/PhysRevA.90.062123}{Phys. Rev. A \textbf{90}, 6} (2014)

\bibitem{Devittetal2013}S. J. Devitt, W. J. Munro, and K. Nemoto,
Quantum error correction for beginners.
\href{https://doi.org/10.1088/0034-4885/76/7/076001}{Rep. Prog. Phys. \textbf{76}, 076001} (2013).

\bibitem{Reineretal2018} J.-M. Reiner, S. Zanker, I. Schwenk, J. Lepp\"akangas, F. Wilhelm-Mauch, G. Sch\"on, and M. Marthaler,
Effects of gate errors in digital quantum simulations of fermionic systems.
\href{https://doi.org/10.1088/2058-9565/aad5ba}{Quantum Sci. Technol. \textbf{3}, 045008} (2018).

\bibitem{NielsenChuang}M.~A.~Nielsen and I.~L.~Chuang,
\textit{Quantum Computation and Quantum Information Ch. 8}.
(Cambridge University Press, 2nd edition, Cambridge 2010).

\bibitem{TrieuThesis}D. B. Trieu,
Large-scale simulations of error-prone quantum computation devices.
Doctoral thesis, Universit\"at  Wuppertal (2009),
published in
\href{http://hdl.handle.net/2128/3688}{Schriften des Forschungszentrums J\"ulich : IAS Series \textbf{2}}.

\bibitem{Chowetal2011}
Jerry M. Chow, A. D. C\'orcoles, Jay M. Gambetta, Chad Rigetti, B. R. Johnson, John A. Smolin, J. R. Rozen, George A. Keefe, Mary B. Rothwell, Mark B. Ketchen, and M. Steffen,
Simple All-Microwave Entangling Gate for Fixed-Frequency Superconducting Qubits,
\href{https://doi.org/10.1103/PhysRevLett.107.080502}{Phys. Rev. Lett. \textbf{107}, 080502} (2011).

\bibitem{Huang}W. Huang, C. H. Yang, K. W. Chan, T. Tanttu, B. Hensen, R. C. C. Leon, M. A. Fogarty, J. C. C. Hwang, F. E. Hudson, K. M. Itoh, A. Morello, A. Laucht, and  A. S. Dzurak,
Fidelity benchmarks for two-qubit gates in silicon.
\href{https://doi.org/10.1038/s41586-019-1197-0}{Nature (London) \textbf{569}, 532} (2019).

\bibitem{EisertPlenio}J.~Eisert and M.~B.~Plenio, A comparison of entanglement measures. \href{https://doi.org/10.1080/09500349908231260}{Journal of Modern Optics \textbf{46}, 145} (1999).

\bibitem{FMezzadri2007}F.~Mezzadri, How to generate random matrices from the classical compact groups.
\href{https://arxiv.org/abs/math-ph/0609050}{https://arxiv.org/abs/math-ph/0609050}

\bibitem{KrausCirac2001}
B. Kraus and J. I. Cirac,
Optimal creation of entanglement using a two-qubit gate,
\href{https://doi.org/10.1103/PhysRevA.63.062309}{Phys. Rev. A \textbf{63}, 062309} (2001).

\bibitem{Khanejaetal2001}
N. Khaneja, R. Brockett, and S. J. Glaser,
Time optimal control in spin systems,
\href{https://doi.org/10.1103/PhysRevA.63.032308}{Phys. Rev. A \textbf{63}, 032308} (2001).

\bibitem{Zhangetal2003}
Jun Zhang, Jiri Vala, Shankar Sastry, and K. Birgitta Whaley,
Geometric theory of nonlocal two-qubit operations,
\href{https://doi.org/10.1103/PhysRevA.67.042313}{Phys. Rev. A \textbf{67}, 042313} (2003).

\bibitem{Fanetal2005}H.~Fan, V.~Roychowdhury, and T.~Szkopek,
Optimal two-qubit quantum circuits using exchange interactions.
\href{https://doi.org/10.1103/PhysRevA.72.052323}{Phys. Rev. A \textbf{72}, 052323} (2005). 

\bibitem{Pedersenetal}L.~H. Pedersen, N.~M.~M\o{}ller and K.~M\o{}lmer,
Fidelity of quantum operations,
\href{https://doi.org/10.1016/j.physleta.2007.02.069}{Physics Letters A \textbf{367}, 47} (2007).

%\bibitem{kus1994random} K.~\.Zyczkowski, M.~Ku\'s, Random unitary matrices.
%\href{}{Journal of Physics A: Mathematical and General \textbf{27}(12), 4235}(1994).

\bibitem{Steffenetal2006}M.~Steffen, M.~Ansmann, R.~C.~Bialczak, N.~Katz, E.~Lucero, R.~McDermott, M.~Neeley, E.~M.~Weig, A.~N.~Cleland, and J.~M.~Martinis,
Measurement of the Entanglement of Two Superconducting Qubits via State Tomography.
\href{https://doi.org/10.1126/science.1130886}{Science \textbf{313}, 1423} (2006).

\bibitem{Zajacetal2018}D.~M.~Zajac, A.~J.~Sigillito, M.~Russ, F.~Borjans, J.~M.~Taylor, G.~Burkard, and J.~R.~Petta,
Resonantly driven CNOT gate for electron spins.
\href{https://doi.org/10.1126/science.aao5965}{Science \textbf{359}, 439} (2018).

\bibitem{Watsonetal2018}T.~F.~Watson, S.~G.~J.~Philips, E.~Kawakami, D.~R.~Ward, P.~Scarlino, M.~Veldhorst, D.~E.~Savage, M.~G.~Lagally, M.~Friesen, S.~N.~Coppersmith, M.~A.~Eriksson, and L.~M.~K.~Vandersypen,
A programmable two-qubit quantum processor in silicon.
\href{https://doi.org/10.1038/nature25766}{Nature \textbf{555}, 633} (2018).

\bibitem{jaccard}P.~Jaccard,
The distribution of the flora in the alpine zone 1.  \href{https://doi:10.1111/j.1469-8137.1912.tb05611.x. ISSN 0028-646X}{ New Phytologist. \textbf{11} (2)  37–50} (1912)
 
\bibitem{Klappenecker2003}A.~Klappenecker and M.~R{\"o}tteler,
Constructions of mutually unbiased bases.
\href{https://doi.org/10.1007/978-3-540-24633-6_10}{Proceedings of the 7th International Conference on Finite Fields and Applications, 137} (2004).

\bibitem{qiskit1}\href{https://qiskit.org/documentation/_modules/qiskit/utils/mitigation/fitters.html}{Qiskit utils mitigation fitters package}

\bibitem{Palmerietal2020}A.~M.~Palmieri, E.~Kovlakov, F.~Bianchi, D.~Yudin, S.~Straupe, J.~D.~Biamonte, and S.~Kulik,
Experimental neural network enhanced quantum tomography.
\href{https://doi.org/10.1038/s41534-020-0248-6}{npj Quantum Information \textbf{6}, 20} (2020). 

\bibitem{SupplMat}Supplemental Material: Code and Data, available under
\href{https://doi.org/10.5281/zenodo.6337555}{https://doi.org/10.5281/zenodo.6337555}

\end{thebibliography}
\end{document}